\documentclass[10pt, twocolumn, tighten]{aastex62}

\usepackage{graphicx}
\usepackage{hyperref}

\newcommand{\hi}{H{\sc i}}
\newcommand{\lya}{Ly$\alpha$}
\newcommand{\lyb}{Ly$\beta$}
\newcommand{\MM}{MAMMOTH-1}
\newcommand{\MQ}{MAMMOTH1-QSO}
\newcommand{\bkg}{$z=2.4$--$2.9$}
\newcommand{\hmpc}{$h^{-1}$ cMpc}

\begin{document}

\title{
3D Distribution Map of \hi\ Gas and Galaxies around an Enormous Ly$\alpha$ Nebula and Three QSOs at $z=2.3$\\
Revealed by the \hi\ Tomographic Mapping Technique
}
\shorttitle{
3D \hi\ Gas Distribution Map Around an ELAN and QSOs
}
\accepted{for Publication in ApJ}

\author{Shiro Mukae}
\affiliation{
Institute for Cosmic Ray Research, The University of Tokyo, 
5-1-5 Kashiwanoha, Kashiwa, Chiba 277-8582, Japan}
\affiliation{
Department of Astronomy, Graduate School of Science, The University of Tokyo, 
7-3-1 Hongo, Bunkyo, Tokyo, 113-0033, Japan}
\email{mukae@icrr.u-tokyo.ac.jp}
\author{Masami Ouchi}
\affiliation{
Division of Science, National Astronomical Observatory of Japan, 2-21-1 Osawa, Mitaka, Tokyo 181-8588, Japan}
\affiliation{
Institute for Cosmic Ray Research, The University of Tokyo, 
5-1-5 Kashiwanoha, Kashiwa, Chiba 277-8582, Japan}
\affiliation{
Kavli Institute for the Physics and Mathematics of the Universe 
(Kavli IPMU, WPI), University of Tokyo, 
5-1-5 Kashiwanoha, Kashiwa, Chiba, 277-8583, Japan}
\author{Zheng Cai}
\affiliation{
Department of Astronomys, Tsinghua University, Beijing 100084, China; zcai@mail.tsinghua.edu.cn}
\author{Khee-Gan Lee}
\affiliation{
Kavli Institute for the Physics and Mathematics of the Universe 
(Kavli IPMU, WPI), University of Tokyo, 
5-1-5 Kashiwanoha, Kashiwa, Chiba, 277-8583, Japan}
\author{J. Xavier Prochaska}
\affiliation{
UCO/Lick Observatory, University of California, 
1156 High Street, Santa Cruz, CA 95064, USA}
\affiliation{
Kavli Institute for the Physics and Mathematics of the Universe 
(Kavli IPMU, WPI), University of Tokyo, 
5-1-5 Kashiwanoha, Kashiwa, Chiba, 277-8583, Japan}
\author{Sebastiano Cantalupo}
\affiliation{
Department of Physics, ETH Zurich, Wolgang-Pauli-Strasse 27, 8093, Zurich}
\author{Yoshiaki Ono}
\affiliation{
Institute for Cosmic Ray Research, The University of Tokyo, 5-1-5 Kashiwanoha, Kashiwa, Chiba 277-8582, Japan}
\author{Zheng Zheng}
\affiliation{
Department of Physics and Astronomy, University of Utah, 115 South
1400 East \#201, Salt Lake City, UT 84112}
\author{Kentaro Nagamine}
\affiliation{
Department of Earth and Space Science, Graduate School of Science, Osaka University, Toyonaka, Osaka 560-0043, Japan}
\affiliation{
Department of Physics \& Astronomy, University of Nevada, Las Vegas, 4505 S. Maryland Pkwy, Las Vegas, NV 89154-4002, USA}
\affiliation{
Kavli Institute for the Physics and Mathematics of the Universe 
(Kavli IPMU, WPI), University of Tokyo, 
5-1-5 Kashiwanoha, Kashiwa, Chiba, 277-8583, Japan}
\author{Nao Suzuki}
\affiliation{
Kavli Institute for the Physics and Mathematics of the Universe 
(Kavli IPMU, WPI), University of Tokyo, 
5-1-5 Kashiwanoha, Kashiwa, Chiba, 277-8583, Japan}
\author{John D. Silverman}
\affiliation{
Kavli Institute for the Physics and Mathematics of the Universe 
(Kavli IPMU, WPI), University of Tokyo, 
5-1-5 Kashiwanoha, Kashiwa, Chiba, 277-8583, Japan}
\author{Toru Misawa}
\affiliation{
School of General Education, Shinshu University, 3-1-1 Asahi, Matsumoto, Nagano 390-8621, Japan}
\author{Akio K. Inoue}
\affiliation{
Department of Physics, School of Advanced Science and Engineering, Waseda University, 3-4-1, Okubo, Shinjuku, Tokyo 169-8555, Japan}
\affiliation{
Waseda Research Institute for Science and Engineering, 3-4-1, Okubo, Shinjuku, Tokyo 169-8555, Japan}
\author{Joseph F. Hennawi}
\affiliation{Department of Physics, University of California, Santa Barbara, 
CA 93106, USA}
\author{Yuichi Matsuda}
\affiliation{
National Astronomical Observatory of Japan, Osawa 2-21-1, Mitaka, Tokyo 181-8588, Japan}
\affiliation{
Department of Genetics, The Graduate University for Advanced Studies, SOKENDAI, Mishima,411-8540 Japan.}
\author{Ken Mawatari}
\affiliation{
Institute for Cosmic Ray Research, The University of Tokyo, 5-1-5 Kashiwanoha, Kashiwa, Chiba 277-8582, Japan}
\author{Yuma Sugahara}
\affiliation{
Institute for Cosmic Ray Research, The University of Tokyo, 5-1-5 Kashiwanoha, Kashiwa, Chiba 277-8582, Japan}
\affiliation{
Department of Physics, Graduate School of Science, The University of Tokyo, 
7-3-1 Hongo, Bunkyo, Tokyo, 113-0033, Japan}
\author{Takashi Kojima}
\affiliation{
Institute for Cosmic Ray Research, The University of Tokyo, 5-1-5 Kashiwanoha, Kashiwa, Chiba 277-8582, Japan}
\affiliation{
Department of Physics, Graduate School of Science, The University of Tokyo, 
7-3-1 Hongo, Bunkyo, Tokyo, 113-0033, Japan}
\author{Takatoshi Shibuya}
\affiliation{
Kitami Institute of Technology, 
165 Koen-cho, Kitami, Hokkaido 090-8507, Japan}
\author{Yuichi Harikane}
\affiliation{
National Astronomical Observatory of Japan, Osawa 2-21-1, Mitaka, Tokyo 181-8588, Japan}
\author{Seiji Fujimoto}
\affiliation{
Department of Physics, School of Advanced Science and Engineering, Waseda University, 3-4-1, Okubo, Shinjuku, Tokyo 169-8555, Japan}
\author{Yi-Kuan Chiang}
\affiliation{
Department of Physics \& Astronomy, Johns Hopkins University, 
3400 N. Charles Street, Baltimore, MD 21218, USA}
\author{Haibin Zhang}
\affiliation{
Institute for Cosmic Ray Research, The University of Tokyo, 5-1-5 Kashiwanoha, Kashiwa, Chiba 277-8582, Japan}
\affiliation{
Department of Physics, Graduate School of Science, The University of Tokyo, 
7-3-1 Hongo, Bunkyo, Tokyo, 113-0033, Japan}
\author{Ryota Kakuma}
\affiliation{
Institute for Cosmic Ray Research, The University of Tokyo, 5-1-5 Kashiwanoha, Kashiwa, Chiba 277-8582, Japan}
\affiliation{
Department of Astronomy, Graduate School of Science, The University of Tokyo, 
7-3-1 Hongo, Bunkyo, Tokyo, 113-0033, Japan}

\begin{abstract}
We present an IGM \hi\ tomographic map in a survey volume of 
$16 \times 19 \times 131 \ h^{-3} {\rm  comoving \  Mpc}^{3}$ (cMpc$^3$) 
centered at \MM\ nebula 
and three neighboring quasars at $z=2.3$. 
\MM\ nebula is an enormous Ly$\alpha$ nebula (ELAN), hosted by a type-II quasar dubbed \MQ, that extends over $1\ h^{-1}$ cMpc with not fully clear physical origin.
Here we investigate the \hi-gas distribution around \MQ\ with the ELAN and 
three neighboring type-I quasars,  
making the IGM \hi\ tomographic map with a spatial resolution of $2.6\ h^{-1}$ cMpc. 
Our \hi\ tomographic map is reconstructed with \hi\ Ly$\alpha$ forest absorption
of bright background objects at $z=2.4-2.9$: one eBOSS quasar and 16 Keck/LRIS galaxy spectra. 
We estimate the radial profile of \hi\ overdensity for \MQ, 
and find that \MQ\ resides in a volume with fairly weak \hi\ absorption. 
This suggests that MAMMOTH1-QSO may have a proximity zone 
where quasar illuminates and photo-ionizes the surrounding \hi\ gas and suppresses \hi\ absorption,
and that the ELAN is probably a photo-ionized cloud embedded in the cosmic web.
The \hi\ radial profile of \MQ\ is very similar to those of 
three neighboring type-I quasars at $z=2.3$, 
which is compatible with the AGN unification model.
We compare the distributions of the \hi\ absorption and star-forming galaxies in our survey volume, 
and identify a spatial offset between density peaks of star-forming galaxies and \hi\ gas. 
This segregation may suggest anisotropic UV background radiation created by star-forming galaxy density fluctuations. 
\end{abstract}


\section{Introduction} \label{sec:introduction}
\par
Enormous \lya\ Nebulae (ELANe) are extremely extended \lya\ nebulae discovered around $z\sim2$ radio-quiet quasars \citep[e.g.,][]{Cantalupo2014a, Kikuta2019, Cai2019a}. 
Since their \lya\ emission extends to $> 1\ h^{-1}$ comoving Mpc (cMpc) beyond the virial diameters of their host quasars,  
the major origin of ELANe is predicted to be quasar photo-ionization of neutral hydrogen (\hi) gas 
embedded in the cosmic web \citep[e.g.,][]{Cantalupo2012a}.
However, the gas distribution around ELANe has been poorly investigated in previous studies.  
\par 
Recently, one of the largest ELANe, \MM, has been identified at $z=2.32$ by \cite{Cai2017a}. 
\MM\ hosts a type-II quasar (hereafter \MQ)  
and, interestingly, resides in a \lya\ emitter (LAE) overdense region 
that has been originally discovered 
with a strong \hi\ 
absorber group, dubbed BOSS1441,  
with multiple background quasar spectra 
in the Mapping the Most Massive Overdensities through Hydrogen (MAMMOTH) survey \citep{Cai2017b}.  
Although 
\cite{Cai2017b} have revealed the existence of the significant \hi\ overdensity around \MM\ 
on a large scale of $\sim 20\ h^{-1}$ cMpc, 
it is yet unknown whether \MQ\ photo-ionizes the surrounding \hi\ gas or not on smaller scales. 
\par 
To investigate the \hi\ gas distribution on such small scales, 
one can use galaxies instead of quasars as background sources 
thanks to their higher number densities 
\citep[e.g.,][]{Mawatari2017a, Hayashino2019a}. 
For this purpose, 
a powerful technique called \hi\ tomography 
has been established by \cite{Lee2014a, Lee2014b}.  
This technique allows us to reconstruct three-dimensional (3D) \hi\ large-scale structures (LSSs) 
based on \hi\ \lya\ forest absorption lines detected in background source spectra 
(See also, \citealt{Lee2016a, Lee2018a}). 
\par 
In this study, we map out the \hi\ gas distribution around \MQ\ with the \hi\ tomography technique 
based on Ly$\alpha$ forest absorption probed with background quasar and galaxy spectra.  
In addition, by combining results of Ly$\alpha$ forest absorption analyses for background quasars 
whose projected distances from \MQ\ 
are relatively large, up to $\sim 200 h^{-1}$ cMpc, 
we estimate the \hi\ radial profile of \MQ\ over a wide range of scales 
and make comparisons with quasars at similar redshifts as well as 
the LAE overdensity distribution. 
\par 
This paper is organized as follows. 
In Section \ref{sec: data and target},
we introduce our background source sample 
and describe their spectroscopic data. 
In Section \ref{sec: analysis}, 
our spectral analyses and \hi\ tomographic reconstruction are described. 
We present results and discussions in Section \ref{sec: results and discussion}.
Finally, we summarize our findings in Section \ref{sec:conclusion}.
Throughout this paper, we use a cosmological parameter set
$( \Omega_m, \Omega_\Lambda, \Omega_b, h)$=$( 0.26, 0.74, 0.045, 0.70)$ 
consistent with the nine-year \textit{WMAP} result \citep{Hinshaw2013a}.
We refer to kpc and Mpc in physical (comoving) units as pkpc and pMpc (ckpc and cMpc), respectively.
All magnitudes are in AB magnitudes \citep{Oke1983a}.

\section{Data and Sample} \label{sec: data and target}
To study the \hi\ gas distribution around \MQ\ over a wide range of scales, 
we need spectra of background quasars and galaxies. 
In this section, 
we construct a spectroscopic sample of background quasars and galaxies 
around the sky position of {\MQ}. 
The basic properties of \MQ\ 
are summarized in Table \ref{tab: MQ}. 

\begin{deluxetable*}{cccccccc}[t]
\tablecolumns{6} 
\tablewidth{0pt}
\vspace{-0.5em}
\tablecaption{Properties of MAMMOTH1-QSO\label{tab: MQ}}
\tablehead{
\colhead{Source}&     \colhead{R.A.} &    \colhead{Decl.} &      \colhead{$z_{\rm spec}$} &   \colhead{$V$} &  \colhead{Ref.}\tablenotemark{a} \vspace{-0.5em}   \\ 
\colhead{ }     &  \colhead{(J2000)} &  \colhead{(J2000)} &                   \colhead{ } &   \colhead{(mag)} &    \colhead{ } 
}
\startdata 
\MQ\ &  14\,41\,24.46\tablenotemark{b} &  +40\,03\,09.20\tablenotemark{b} &  $2.319$ &  $24.20$ & C17
\enddata 
\centering
\tablenotetext{$a$}{
C17: \cite{Cai2017a} }
\tablenotetext{$b$}{
Updated coordinates in Keck/KCWI observations of Z. Cai et al. (in prep.) }
\end{deluxetable*}

\subsection{Background Quasars} \label{sec: background quasar sample}
We select background quasars around {\MQ} 
from the SDSS DR14 quasar catalog \citep[hereafter DR14Q;][]{Paris2018a},
which includes all quasars identified by the SDSS-IV/eBOSS survey \citep{Myers2015a}.
The eBOSS spectra have a spectral resolution of $R \equiv \lambda / \Delta \lambda \approx 2000$    
covering a wavelength range of $3600$--$10400${\AA}, 
which is sufficient for our purpose.   

First, we search for quasars in a $6\fdg0 \times 6\fdg0$ field 
around \MQ\ (the $6\fdg0$ scale corresponds to a span of 400 \hmpc). 
We then select quasars whose emission redshifts are in the range of $z=2.4$--$2.9$ 
so that we can probe Ly$\alpha$ forest absorption lines at the redshift of \MQ\ 
in the background quasars' rest-frame $1041$--$1185${\AA} spectral region 
to avoid contamination of 
\hi\ \lyb\ absorption and stellar/interstellar absorption lines 
associated with the quasar host galaxies 
\citep[e.g.,][]{Mukae2017a}.  
These two criteria yield $240$ DR14Q quasars. 

To obtain robust measurements of Ly$\alpha$ forest absorption,  
we check the qualities of the eBOSS spectra 
and remove quasars whose spectra do not meet additional criteria described below.  
We require the eBOSS spectra have a median signal-to-noise ratio (S/N) $\geq 2$  
over their \lya\ forest wavelength range (i.e., $1041$--$1185${\AA} in the rest-frame). 
In addition, we remove quasars whose spectra have broad absorption lines
by applying BI $<200$ km s$^{-1}$ in the DR14Q catalog, 
where BI (BALnicity Index) is 
a measure of the strength of an absorption trough calculated for the {\sc Civ} emission line. 
We also remove quasars whose spectra show damped \lya\ systems (DLAs) 
in the \lya\ forest wavelength range, 
based on the DLA catalog of \cite{Noterdaeme2012a} 
and their updated one\footnote{ \url{http://www2.iap.fr/users/noterdae/DLA/DLA.html} }
for the SDSS DR12 quasars \citep{Paris2017a}.
For quasars that have no SDSS DR12 counterpart, 
we visually inspect the eBOSS spectra and remove them 
if they show signatures of DLAs in their \lya\ forest wavelength range.
Our careful selection results in a sample of $117$ background quasars 
for our subsequent analyses. 

\begin{figure*}[t]
 {\centering
 \includegraphics[width=0.80\hsize, clip, bb=0 0 900 700, clip=true]{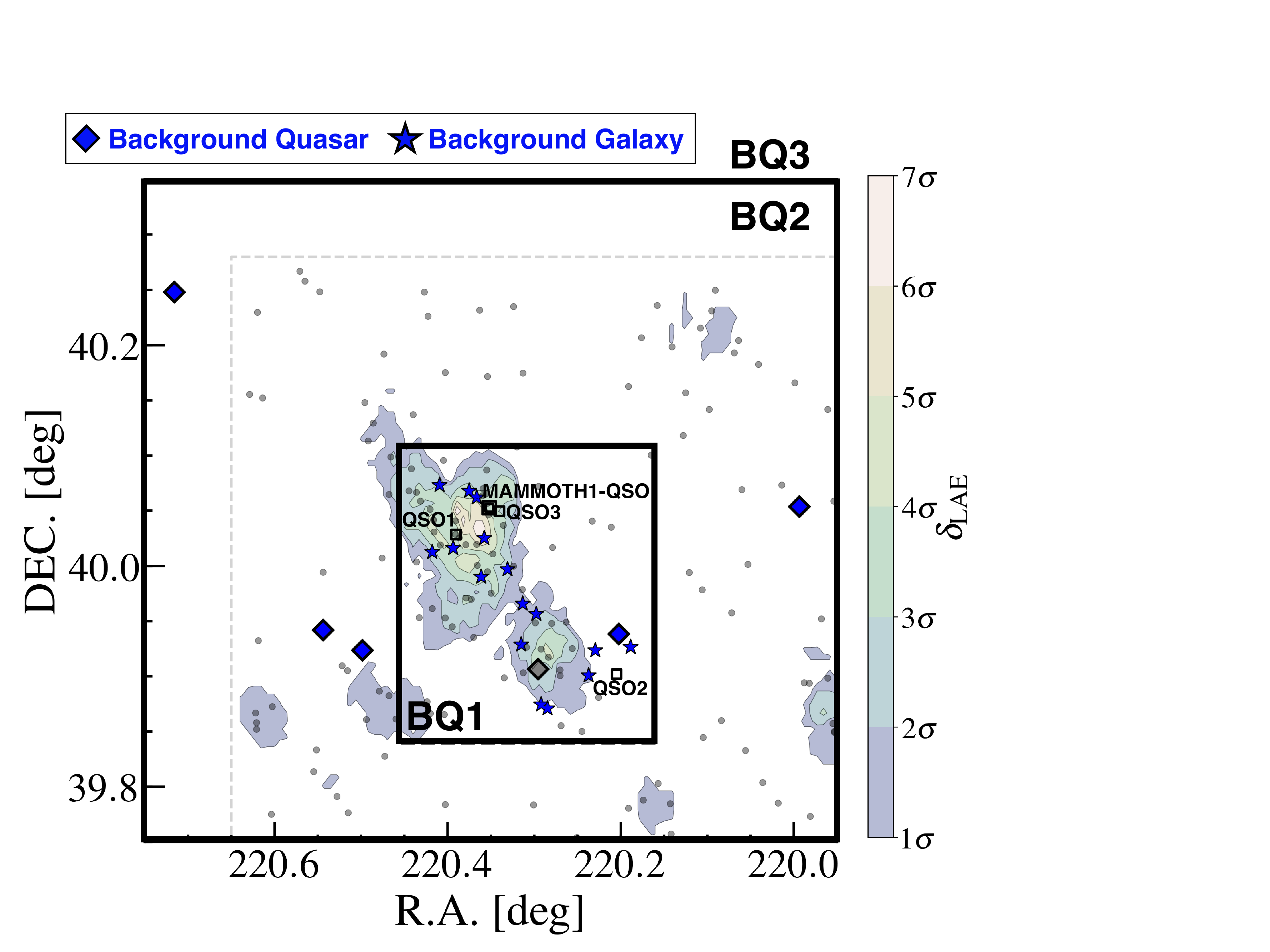}
 \caption{
 Sky distribution of our background quasars and galaxies  in the BQ1 and BQ2 regions. 
 The blue diamonds (stars) represent the positions of our background quasars (galaxies) at \bkg. 
 The gray diamond indicates a $z > 2.9$ quasar that is used in \cite{Cai2017b} in addition to the other five quasars, 
 although we do not use it in our analyses (for details, see Section \ref{sec: background quasar sample}).
 The double square is the position of \MQ\ at $z=2.32$ and 
 the single squares are those of three neighboring quasars at $z \simeq 2.3$, QSOs 1--3 (Section \ref{sec: qpq6}).
 The dark-gray dots are $z \simeq 2.32$ LAEs found in the previous work (\citealt{Cai2017b}), 
 whose survey area is shown with light-gray dashed lines. 
 The color contours represent LAE overdensities $\delta_{\rm LAE}$ calculated in Section \ref{sec: LAE-HI overdensity}.
 }
 }
 \label{fig: sky}
 \end{figure*}

For convenience, 
we divide the $6\fdg0 \times 6\fdg0$ field around \MQ\ into three regions, 
BQ1, BQ2, and BQ3 as illustrated in Figure \ref{fig: sky}. 
The boundary between BQ1 and BQ2 is defined with a rectangle whose corners are 
($\Delta$R.A., $\Delta$Decl.) = 
($-0\fdg19$, $-0\fdg21$), ($-0\fdg19$, $+0\fdg058$), ($+0\fdg11$, $+0\fdg058$), and ($+0\fdg11$, $-0\fdg21$)
relative to the coordinate of {\MQ},
so that the 16 \hmpc\ $\times$ 19 \hmpc\ area of the BQ1 region 
can cover the position of \MQ\ 
as well as the LAE overdense region for comparison in Section \ref{sec: LAE-HI overdensity}. 
The boundary between BQ2 and BQ3 is defined with a rectangle whose corners are 
$\Delta$R.A. = $\pm 0\fdg4$ and $\Delta$Decl. = $\pm 0\fdg3$. 
This corresponds to a 40 \hmpc\ $\times$ 41 \hmpc\ rectangle
atop the entire BOSS1441 field \citep{Cai2017b}.
In the BQ1, BQ2, and BQ3 regions,  
the numbers of our background  quasars are 1, 4, and 112, respectively. 
The distributions of the background quasars in BQ1 and BQ2 regions are presented in Figure \ref{fig: sky}.
The basic properties of the background quasars in the BQ1 and BQ2 regions
are provided in Tables \ref{tab: bgsl} and \ref{tab: bq boss bq2}, respectively.

Note that \cite{Cai2017b} have identified the strong \hi\ absorber group 
based on the spectra of these five background quasars as well as an additional one in the BQ1--2 regions. 
In our analyses we do not use the spectrum of the additional quasar, 
since its redshift is $z>2.9$ and  
there is a possibility that the spectrum is contaminated by the Ly$\beta$ absorption 
and/or stellar/interstellar absorption due to the quasar host galaxy. 

\begin{deluxetable*}{ccccccccc}[ht]
\tablecolumns{6}
\tablewidth{0pt}
\vspace{-0.3em}
\tablecaption{Background objects for  \hi\ tomographic reconstruction in the BQ1 region 
\label{tab: bgsl} }
\tablehead{
\colhead{Source}\vspace{-0.0em} & 
\colhead{R.A.} & 
\colhead{Decl.} & 
\colhead{$z_{\rm spec}$} & 
\colhead{$g$} & 
\colhead{$V$} & 
\colhead{Exposure Time} &
\colhead{Sample\tablenotemark{a}}
& \vspace{-0.5em} \\ 
\colhead{ }\vspace{-0.0em}  & 
\colhead{(J2000)} & 
\colhead{(J2000)} & 
\colhead{ } & 
\colhead{(AB)} & 
\colhead{(AB)} & 
\colhead{(s)} & 
\colhead{} 
} \vspace{-1em}
\startdata 
BQ1-5172-56071-0534 & 14\,40\,48.56 & +39\,56\,18.39 & 2.543 & 20.04 & - & - & eBOSS \\ 
20170827\_M1\_05  & 14\,41\,19.44 & +39\,59\,49.52 & 2.509 & - & 24.04 & 9000 & LRISs \\ 
20170827\_M1\_07  & 14\,41\,26.77 & +39\,59\,25.01 & 2.816 & - & 23.40 & 9000 & LRISs \\ 
20170827\_M1\_22  & 14\,41\,27.98 & +40\,03\,43.31 & 2.510 & - & 24.50 & 9000 & LRISs \\ 
20170827\_M1\_24  & 14\,41\,30.05 & +40\,04\,05.59 & 2.671 & - & 23.58 & 9000 & LRISs \\ 
20160510\_M2\_05  & 14\,41\,11.50 & +39\,57\,24.08 & 2.546 & - & 24.11 & 5400 & LRISa \\ 
20160510\_M2\_10  & 14\,41\,16.63 & +39\,58\,51.56 & 2.598 & - & 23.21 & 5400 & LRISa \\
20160510\_M2\_25  & 14\,41\,34.49 & +40\,00\,58.68 & 2.795 & - & 23.61 & 5400 & LRISa \\
20160509\_M1\_11  & 14\,41\,40.29 & +40\,00\,46.08 & 2.557 & - & 23.05 & 4000 & LRISa \\ 
20160509\_M1\_23  & 14\,41\,38.31 & +40\,04\,23.49 & 2.786 & - & 22.84 & 4000 & LRISa \\
20160405\_M1\_05  & 14\,41\,25.85 & +40\,01\,31.40 & 2.795 & - & 22.82 & 6000 & LRISa \\ 
20160405\_M2\_08  & 14\,41\,10.15 & +39\,52\,28.88 & 2.791 & - & 24.27 & 6000 & LRISa \\ 
20160405\_M2\_10  & 14\,41\,15.62 & +39\,55\,43.97 & 2.512 & - & 24.42 & 6000 & LRISa \\ 
20160405\_M2\_11  & 14\,41\,08.38 & +39\,52\,15.59 & 2.497 & - & 23.68 & 6000 & LRISa \\ 
20160405\_M2\_20  & 14\,40\,56.98 & +39\,54\,03.24 & 2.703 & - & 24.34 & 6000 & LRISa \\ 
20160405\_M2\_27  & 14\,40\,55.13 & +39\,55\,25.70 & 2.840 & - & 24.37 & 6000 & LRISa \\ 
20160405\_M2\_35  & 14\,40\,45.31 & +39\,55\,35.76 & 2.598 & - & 23.07 & 6000 & LRISa 
\enddata 
\centering
\tablenotetext{$a$}{
eBOSS: eBOSS spectroscopy data  (Section \ref{sec: background quasar sample}), \\
LRISs: Keck spectroscopy data (Section \ref{sec: keck spectroscopy}), LRISa: Keck archive data (Section \ref{sec: keck archive}) }
\end{deluxetable*} 

\begin{deluxetable*}{cccccccc}[t]
\tablecolumns{6} 
\tablewidth{0pt}
\vspace{-0em}
\tablecaption{
Background quasars in the BQ2 region
\label{tab: bq boss bq2} }
\tablehead{
\colhead{Source}&     \colhead{R.A.} &    \colhead{Decl.} &      \colhead{$z_{\rm spec}$} &  \colhead{$g$} &   \colhead{Ref.}\tablenotemark{a}   \\ 
\colhead{ }     &  \colhead{(J2000)} &  \colhead{(J2000)} &                   \colhead{ } & \colhead{(AB)} &     \colhead{ }
}
\startdata 
BQ2-5171-56038-0020 &  14\,39\,58.45  & +40\,03\,13.99 & 2.422 & 20.20 &  DR14Q\\ 
BQ2-8498-57105-0478 &  14\,41\,59.76  & +39\,55\,25.32 & 2.546 & 19.47 &  DR14Q\\ 
BQ2-5172-56071-0616 &  14\,42\,10.56  & +39\,56\,31.92 & 2.612 & 20.99 &  DR14Q\\ 
BQ2-5172-56071-0608 &  14\,42\,51.84  & +40\,14\,53.52 & 2.547 & 20.86 &  DR14Q\\ 
\enddata 
\centering
\tablenotetext{$a$}{DR14Q: \cite{Paris2018a} }
\end{deluxetable*}

\subsection{Background Galaxies} \label{sec: background galaxy target}
\subsubsection{Candidate Selection} \label{sec: target for spectroscopy}
As shown in Figure \ref{fig: sky}, 
the number of our background quasars near \MQ\ is small,  
only one in the BQ1 region. 
To investigate the \hi\ gas distribution around \MQ\ down to a small scale, 
we need a sample of close background galaxies at $z = 2.4$--$2.9$ near \MQ. 

For this purpose, 
we produce a multiwavelength catalog across the BQ1 region 
based on optical ($U$, $V$, and $i$) and near-IR ($J$ and $H$) imaging data 
obtained by \cite{Cai2017b} and Z. Cai et al. (in prep.) 
with 
the Large Binocular Camera \citep[LBC;][]{Pedichini2003a}
on the Large Binocular Telescope (LBT)
and the Wide Field Camera \citep[WFCAM;][]{Casali2007a}
on the United Kingdom Infrared Telescope (UKIRT), 
respectively. 
We match the point spread functions (PSFs) of these images 
to that of the $H$-band image 
whose FWHM ($0\farcs9$) is the largest among them.
The $5\sigma$ limiting magnitudes 
in the $U$, $V$, $i$, $J$, and $H$ bands 
measured with $2\farcs0$ diameter apertures 
are $26.6$, $26.2$, $26.1$, $23.7$, and $23.1$ mag, respectively. 
We then create a multiwavelength source catalog 
by running SExtractor \citep{Bertin1996a} in dual image mode. 
Colors of the detected sources are measures with $2\farcs0$ diameter apertures.   

Based on the multiwavelength catalog, 
we estimate photometric redshifts of the detected sources 
to select background galaxy candidates at $z=2.4$--$2.9$. 
First of all, 
we apply a magnitude cut of $V < 24.85$ mag 
so that we can select background galaxy candidates whose continuum emission 
can be detected with sufficiently high S/Ns 
in subsequent spectroscopic observations. 
We then estimate their photometric redshifts 
with the EAZY software \citep{Brammer2008a}  
by fitting spectral energy distribution (SED) templates 
to the observed photometric data points. 
The SED templates are produced 
with the stellar population synthesis model of \cite{Bruzual2003a}.
We adopt the Chabrier initial mass function \citep{Chabrier2003a},
a constant star formation for $0.1$ Gyr,
and a fixed metallicity of $Z=0.2 Z_{\odot}$. 
The metallicity is chosen to be close to the 
metallicity estimates of star-forming galaxies
in \citet{Pettini2000a} and \citet{Pettini2001a}.
We apply the Calzetti dust attenuation \citep{Calzetti2000a}
with $E(B-V) = $ 0.0, 0.15, 0.30, and 0.45.
We also apply attenuation by IGM absorption
with a model of \cite{Inoue2014a}.
We require background galaxy candidates 
to have a photometric redshift 
whose $1\sigma$ confidence interval is within the redshift range of $z=2.4$--$2.9$. 
The $1\sigma$ errors on the photometric redshift from the used software are $\sim0.17$.
This selection yields a sample of $131$ background galaxy candidates in BQ1.

\subsubsection{Follow-up Spectroscopy} \label{sec: keck spectroscopy}
We carried out spectroscopic observations for our background galaxy candidates 
using the Low Resolution Imaging Spectrometer (LRIS) Double-Spectrograph \citep{Oke1995a}
on the Keck I telescope on 2017 August 27 (UT) (PI: S. Mukae).
We used the d560 dichroic with the B600/4000 grism on the blue arm, 
resulting in a wavelength coverage of 3800--5500{\AA}. 
The observations were made in the multi-object slit (MOS) mode. 
We designed one mask 
targeting background galaxy candidates around {\MQ} 
with $1\farcs0$ slit width,  
yielding a spectroscopic resolution of 
$R \equiv \lambda / \Delta \lambda \approx 1000 $. 
We select 25 background galaxies
from our 131 background galaxy candidates
based on photometric redshift probability, source brightness, and uniformity on the sky.
The total exposure time was 9000 s.
The sky conditions were clear throughout the observing run, 
with an average seeing size of $0\farcs95$.

We reduce the LRIS data with the Low-Redux package\footnote{ \url{http://www.ucolick.org/~xavier/LowRedux/} }
in the public XIDL pipeline.\footnote{ \url{http://www.ucolick.org/~xavier/IDL/} } 
The pipeline conducts bias subtraction, 
flat fielding with dome flat and twilight flat data, 
wavelength calibration with arc data, 
cosmic ray rejection, 
source identification, 
spectral trace determination, 
sky background subtraction, 
and distortion correction. 
We then extract one-dimensional (1D) spectra of the identified sources 
from the reduced two-dimensional (2D) spectra 
and combine them to obtain their stacked 1D spectra. 

\subsubsection{Archival Search} \label{sec: keck archive}
In addition to our own observations, 
two other LRIS programs were conducted for the BOSS1441 field 
in the MOS mode 
on 2016 April 5 (UT)  (PI: X. Fan) 
and 2016 May 9--10 (UT) (PI: X. Prochaska) 
by using the same dichroic and grism as ours. 
Although the original aim of their LRIS programs is 
to identify associated galaxies in the BOSS1441 overdense region at $z=2.32$ 
(Z. Cai et al., private communication), 
there is a possibility that 
some background galaxies at $z=2.4-2.9$ are included as targets in the MOS masks 
and spectroscopically identified by chance.  
Thus, we download raw LRIS data 
obtained in the two programs 
from the Keck Observatory Archive (KOA)\footnote{ \url{https://www2.keck.hawaii.edu/koa/public/koa.php} } 
and reduce them in the same manner as for our LRIS data. 

\subsubsection{Sample Construction for Our Analyses} \label{sec: galaxy spec$z$}
We determine spectroscopic redshifts ($z_{\rm spec}$) of the observed sources  
based on the LRIS spectra obtained in Sections \ref{sec: keck spectroscopy} and \ref{sec: keck archive}.  
We fit the galaxy spectrum template of \cite{Shapley2003a} to the LRIS spectra 
and determine the best-fit $z_{\rm spec}$ by the minimum value of $\chi^2$. 
We find that $20$ galaxies have $z_{\rm spec}$ values in the range of $2.4$--$2.9$. 

To obtain robust measurements of Ly$\alpha$ forest absorption, 
in the same way as 
for the background quasars, 
we further require that 
the spectra of background galaxies have a median S/N $\geq 2$ 
in the Ly$\alpha$ forest wavelength range 
of $1041$--$1185${\AA} in the rest-frame.  
In addition, 
based on our visual inspection, 
we remove a galaxy whose spectrum 
shows a possible feature of a DLA 
in the Ly$\alpha$ forest wavelength range. 
These selections result in 
a sample of $16$ background galaxies.  
The basic properties of the $16$ background galaxies 
in the BQ1 region are summarized in Table \ref{tab: bgsl}. 

\begin{figure}[t]
 {\centering
 \includegraphics[width=1.0\hsize, clip, bb= 0 0 650 700, clip=true]{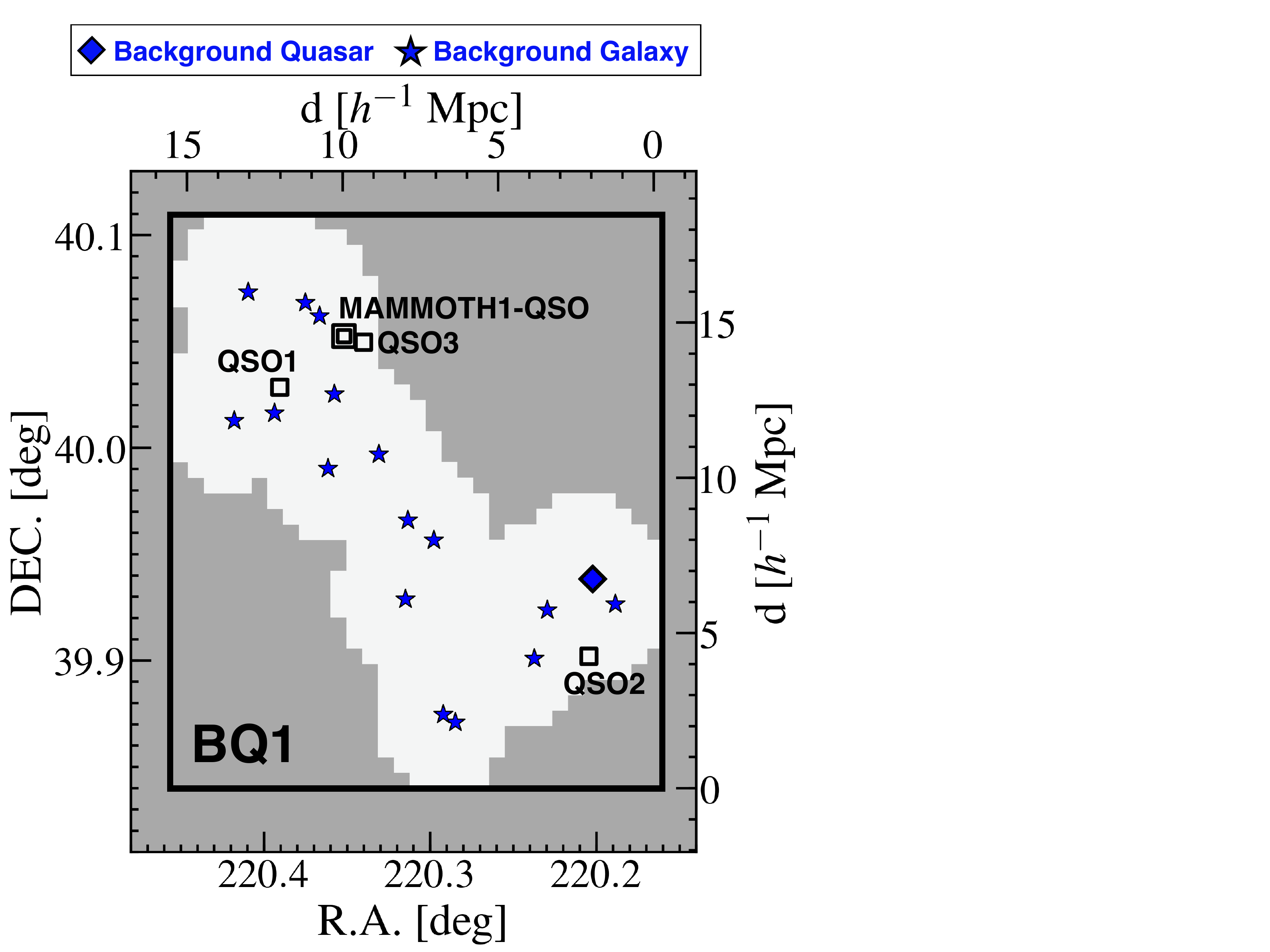}
 \caption{
 Sky distribution of our background sources in the BQ1 region. 
 The blue diamond and stars represent the positions of our background quasar and galaxies, respectively. 
 The white area highlights regions around the sightlines around the background sources 
 within the mean transverse sightline separation ${\langle d}_{\perp} \rangle = 2.6 h^{-1}$ cMpc. 
 The double square denotes the position of \MQ\ and 
 the single squares show the those of three neighboring quasars, QSOs 1--3 (Section \ref{sec: qpq6}).
 The top and right axes are co-moving separations 
 relative to the bottom right corner coordinate of the BQ1 region, 
 (R.A., Decl.) = ($220\fdg16$, $39\fdg84$). 
 }
 }
 \label{fig: bkgls}
\end{figure}

Figure \ref{fig: bkgls} shows 
the positions of the background quasar and galaxies in the BQ1 region. 
The mean (median) transverse sightline separation is ${\langle d}_{\perp} \rangle \simeq 2.6\ (2.7)\ h^{-1}$ cMpc, 
which is comparable to that of the \hi\ tomographic map of \cite{Lee2014b}.
The filling factor of the sky coverage, 
which is defined as the fraction of the regions around the sightlines of the background sources 
within ${\langle d}_{\perp} \rangle$ in the BQ1 region, is about $0.45$. 

\section{Analysis} \label{sec: analysis}
\subsection{Ly$\alpha$ Forest Transmission} \label{sec: continuum estimation}
To measure the strengths of Ly$\alpha$ forest absorption around \MQ\ 
along the lines of sight to the background quasars and galaxies,
we estimate the \lya\ forest transmission, 
\begin{equation}
F (z) = f_{\rm obs} / f_{\rm int}, 
\label{eq:Fz}
\end{equation}
where $z$ is the Ly$\alpha$ absorber redshift 
derived from the observed wavelength $\lambda_{\rm obs}$
(i.e., $\lambda_{\rm obs}/1216$\AA-1), 
$f_{\rm obs}$ is the observed continuum flux density, 
and  $f_{\rm int}$ is the intrinsic continuum flux density 
that is not affected by the Ly$\alpha$ forest absorption due to the IGM.  
The transmission $F (z)$ is computed pixel by pixel 
with a pixel scale of 0.8 (1.2) \AA\ for our eBOSS (LRIS) spectra.

\begin{figure}[t]
 \centering
 \includegraphics[width=1.0\hsize, clip, bb= 0 0 700 800, clip=true]{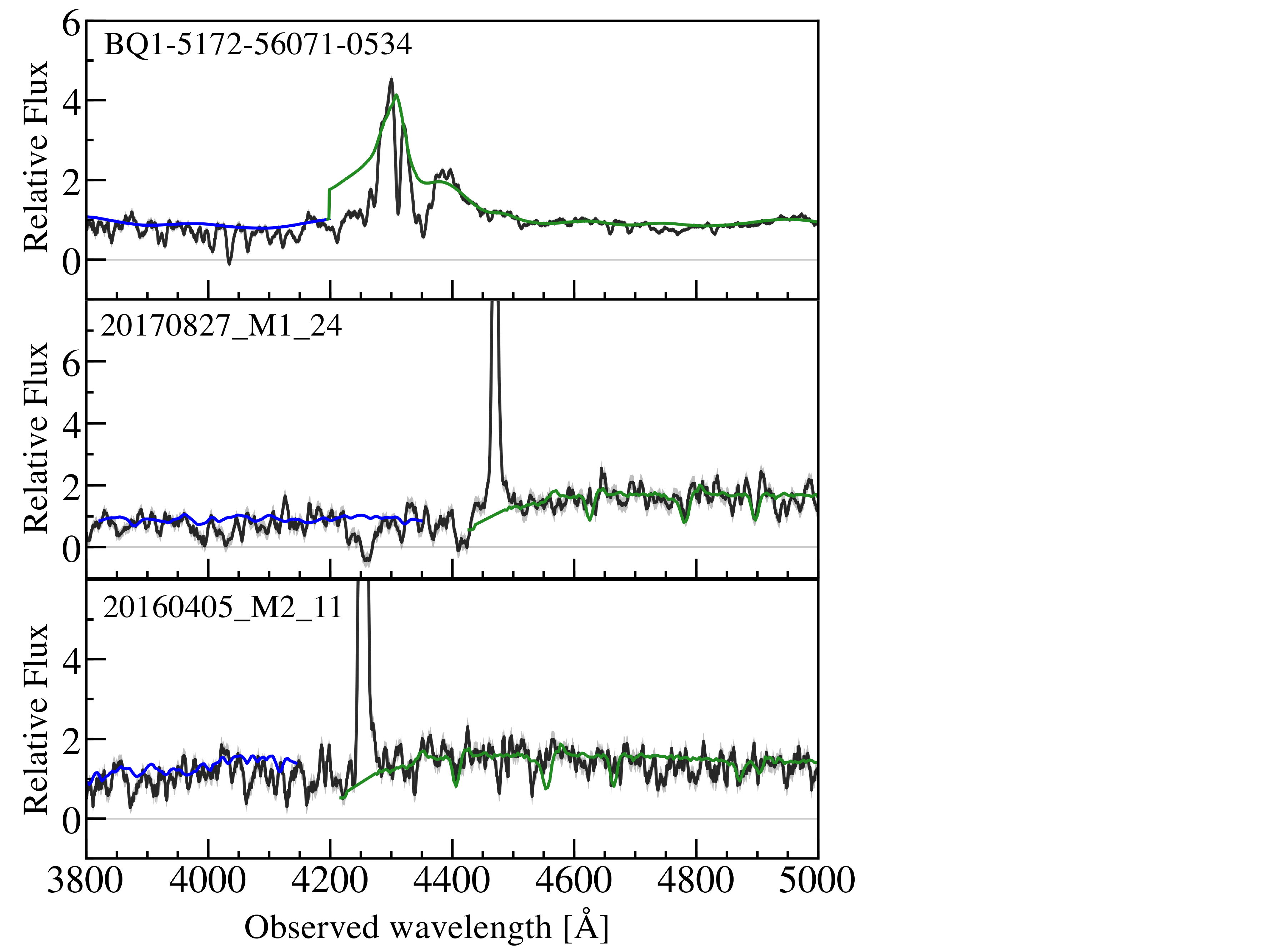}
 \caption{
 Example spectra of our background sources in the BQ1 region. 
 Top panel is an eBOSS spectrum of a background quasar.
 Middle and bottom panels are Keck/LRIS spectra of background galaxies.
 The black lines depict the spectra smoothed with $1.25$ \hmpc\ scales along the sightlines and
 the grey shades represent the uncertainties.
 The grey lines are the flux zero level.
 The green curve(s) in the top (middle and bottom) panel(s) are the \cite{Suzuki2005a}
 (\citealt{Shapley2003a}) template overplotted at the source redshifts.
 The blue curve(s) in the top (middle and bottom) panel(s) are the \cite{Suzuki2005a}
 (\citealt{Berry2012a}) template representing the estimated continuum (continua)
 within the \lya\ forest wavelength range (for details, see Section \ref{sec: continuum estimation}).
 }
 \label{fig: cont}
\end{figure}

We estimate $f_{\rm int}$ 
based on the spectra of the background quasars and galaxies in the BQ1--3 regions 
by applying the 
mean flux regulated principal component analysis 
(MF-PCA) continuum fitting technique \citep{Lee2012a}
with the code developed by \cite{Lee2013a} 
(see also \citealt{Lee2014b}).
This technique is composed of two steps. 
The first step is to fit 
spectral templates of quasars and galaxies 
with the observed spectra in the redward of Ly$\alpha$ 
to obtain initial guesses of their continuum spectra in the blueward of Ly$\alpha$. 
We use the spectral templates of quasars and galaxies 
constructed by \cite{Suzuki2005a} and \cite{Berry2012a}, respectively. 
The second step is to further adjust the continuum spectra
by multiplying and fitting a linear function, ($a_{\rm MF}$ + $b_{\rm MF}$ $\lambda_{\rm rest}$), 
where $a_{\rm MF}$ and $b_{\rm MF}$ are free parameters for the fit, 
and $\lambda_{\rm rest}$ is the rest-frame wavelength. 
This fit is performed for the continuum spectra 
within the \lya\ forest wavelength range of 1041-1185 \AA\ in the rest-frame
to yield a mean transmission (using Equation (\ref{eq:Fz})) consistent 
with previous measurements of the cosmic mean \lya\ forest transmission, ${F_{\rm cos} (z)}$. 
We adopt ${F_{\rm cos} (z)}$ estimated by \cite{Faucher-Giguere2008a}, 
\begin{eqnarray}
	{F_{\rm cos} (z)}=\exp[-0.00185(1+z)^{3.92}],
	\label{eq:mean flux}
\end{eqnarray}
where $z$ is the Ly$\alpha$ absorber redshift.
Both the examples of continuum estimates for 
our background quasar and galaxy spectra are shown in Figure \ref{fig: cont}.
Although the MF-PCA technique introduces a discontinuity at 1185 \AA\ in the final continuum, 
the discontinuity does not affect our analysis in the \lya\ forest wavelength range blueward of 1185 \AA\ (see \citealt{Lee2013a, Lee2012a} for more details). 

There is a possibility that 
the strong \hi\ absorption group at $z=2.32 \pm 0.03$ 
found by \cite{Cai2017b}
could bias the intrinsic continuum estimate. 
To avoid possible contamination 
of their strong Ly$\alpha$ absorption,  
we mask out the wavelength range of $4036\pm36${\AA} 
in the MF-PCA fitting. 

We then obtain $F (z)$ by using Equation (\ref{eq:Fz}).   
Since the strong stellar and interstellar absorption lines 
of N{\sc ii} $\lambda 1084$ and C{\sc iii} $\lambda 1175$ 
associated with 
background sources 
in the Ly$\alpha$ forest wavelength range 
could bias the results, 
we do not use the spectra in the wavelength ranges of $\pm5${\AA} around these lines 
for conservative estimates 
in the following analyses.  
The uncertainties of $F(z)$ are calculated  
from the uncertainties of the $f_{\rm obs}$ measurements and the $f_{\rm int}$ estimates 
based on the MF-PCA continuum fitting, 
the latter of which are evaluated by \cite{Lee2012a} 
as a function of redshift and median S/N over the Ly$\alpha$ forest wavelength range 
(see their Figure 8).  
Specifically, we adopt MF-PCA continuum fitting errors of 
$7${\%}, $6${\%}, and $4${\%} 
for spectra with median S/Ns over the Ly$\alpha$ forest wavelength range of 
$2$--$4$, $4$--$10$, and $>10$, respectively.  

Based on the estimated $F(z)$ and the cosmic mean Ly$\alpha$ forest transmission $F_{\rm cos}(z)$, 
we calculate the \hi\ overdensity $\delta_{F}$, following the definition introduced by \cite{Lee2014a, Lee2014b},
\begin{eqnarray}
    \delta_{F} = \frac{F(z)}{F_{\rm cos} (z)}-1,
    \label{eq: flux overdensity}
\end{eqnarray}
where negative values correspond to strong \hi\ \lya\ absorption. 
The uncertainties of $\delta_F$ are calculated 
based on the uncertainties of $F(z)$. 
We confirm that a systematic effect of using different prescriptions of $F_{\rm cos}(z)$ 
obtained by \cite{Becker2013a} and \cite{Inoue2014a} is minor, only within $2${\%}, 
which is not as large as the uncertainties of $F(z)$. 

\subsection{\hi\ Tomographic Reconstruction} \label{sec: HI Tomography reconstruction}
For the BQ1 region, where the background sightline density is high, 
we carry out an \hi\ tomographic reconstruction 
to reveal the 3D distribution of \hi\ gas near \MQ\ 
with the code developed by \cite{Stark2015a}.\footnote{ \url{https://github.com/caseywstark/dachshund} } 
The reconstruction code performs the Wiener filtering 
for the estimated $\delta_F$ values along the sightlines of our background quasar and galaxies.
The Wiener filtering is based on a gaussian smoothing with the scale of $\langle d_{\perp }\rangle$, 
which determines the spatial resolution of our tomographic map. 
We adopt a grid size of $0.5\ h^{-1}$ cMpc, 
which is sufficiently small compared to ${\langle d}_{\perp} \rangle$. 
We choose a redshift range of $z=2.25$--$2.40$ 
that covers a large distance of $\sim \pm 65\ h^{-1}$ cMpc 
from \MQ\ at $z=2.32$ in the redshift direction,  
giving an overall volume 
of $16\ \times 19\ \times 131\ h^{-3} {\rm cMpc}^{3}$. 
More details about the reconstruction process is presented 
in \cite{Stark2015a} and \cite{Lee2018a}.

There is is a possibility that sightlines used in the \hi\ tomography could undersample 
small \hi\ gas clumps in the \lya\ forest on scales below ${\langle d}_{\perp} \rangle$. 
We thus do not use the small-scale gas distributions in our discussions from Section \ref{sec: HI map}.
The simulation studies of \cite{Lee2014a} demonstrated that
an \hi\ tomography pixel could have the $\delta_F$ error of $\leq 0.05$
due to the sightline undersampling of the \lya\ forest.

Note that we do not reconstruct 
\hi\ tomographic maps for the BQ2-3 regions
due to the coarse sightline distributions 
whose sightline separations are $\sim$15-20 \hmpc. 
Instead, we use the BQ2-3 background quasars 
for large-scale $\delta_F$ measurements along the sightlines (Section \ref{sec: spatial correlation}).

\section{Results and Discussion} \label{sec: results and discussion}
\subsection{\hi\ Tomographic Map} \label{sec: HI map}
\begin{figure*}[t]
 {\centering
 \includegraphics[width=1.0\hsize, clip, bb=0 10 1050 565, clip=true]{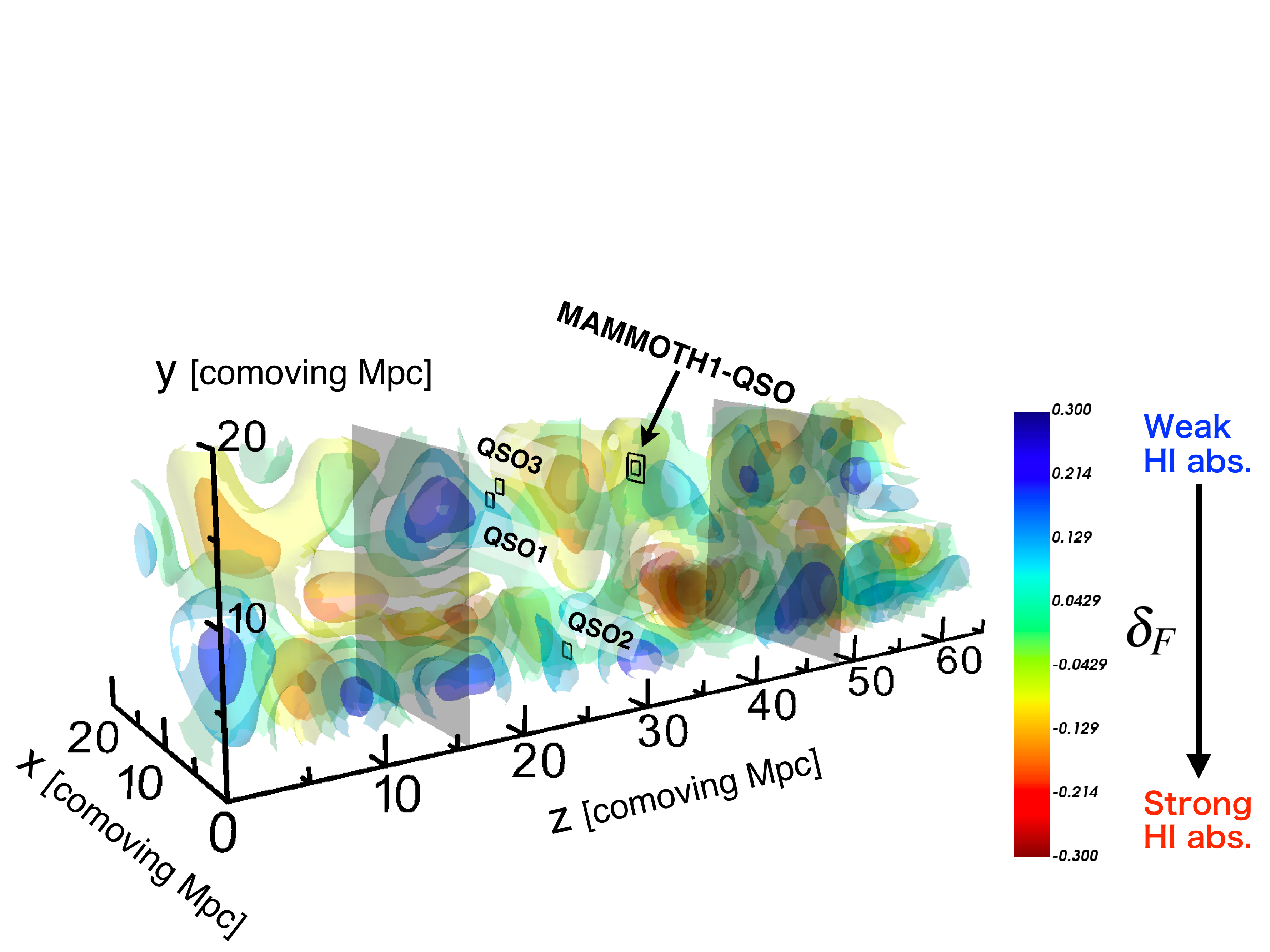}
 \caption{
 \hi\ tomographic map reconstructed based on Ly$\alpha$ forest absorption  
 along the sightlines of our background quasar and galaxies in the BQ1 region.
 Here we present a zoomed map at $z=2.28$--$2.35$ 
 to clearly show the \hi\ overdensity distribution around {\MQ}, 
 although we obtain the \hi\ tomographic map in the redshift range of $z=2.25$--$2.40$. 
 The spatial axes of R.A., Decl., and $z$ are represented as $x$, $y$, and $z$ in co-moving scales, respectively.
 The color contours show the \hi\ overdensity $\delta_{F}$,  
 whose maximum (minimum) scale is set to $+0.3$ ($-0.3$) for visualization, 
 although some regions show higher or lower $\delta_F$ values in the range of $-0.6 < \delta_F < 0.4$. 
 The double square indicates the position of \MQ. 
 The single squares show the three neighboring quasars, QSOs 1--3 (Section \ref{sec: qpq6}).
 Note that the redshift range between the two gray planes at $z \simeq 2.30$ and $z \simeq 2.33$ 
 correspond to the FWHM of the narrowband filter NB403, 
 which is used in \cite{Cai2017a, Cai2017b} to detect 
 LAEs at $z \simeq 2.30$--$2.33$
 in the BOSS1441 field (Section \ref{sec: LAE-HI overdensity}). 
 }
 }
 \label{fig: HI3D}
 \end{figure*}

Figure \ref{fig: HI3D} presents the resulting \hi\ tomographic map 
for the BQ1 region. 
Our tomographic map shows $\delta_{F}$ values in the range of $-0.6 < \delta_F < 0.4$, 
revealing the existence of 
\hi\ overdense ($\delta_{F} \simeq -0.3-(-0.2)$) and underdense  ($\delta_{F} \simeq 0.3-0.2$) LSSs
with sizes of $10$--$20\ h^{-1} {\rm cMpc}$ around \MQ\ for the first time. 
In this region, the strong \hi\ 
absorption group at $z=2.32$ has been found in the previous work 
based on only the six background quasar spectra \citep{Cai2017b}  
as mentioned in Section \ref{sec: background quasar sample}. 
Thanks to the higher sightline density of our background source sample in this field, 
our results 
unveil the inhomogeneous distribution of \hi\ gas around \MQ. 

\subsection{\hi\ Radial Profile} \label{sec: spatial correlation}
\subsubsection{\hi\ gas around \MQ} \label{sec: mq}
In the BOSS1441 region, there is a type-II quasar dubbed \MQ\ 
that has one of the largest ELANe, \MM\ nebula at $z=2.32$ \citep{Cai2017a}.
Since the \lya\ emission spatially extends to $> 1\ h^{-1}$ cMpc beyond virial diameter of the host quasar, 
the origin of \MM\ nebula is thought to be 
quasar photo-ionization of \hi\ gas cloud 
embedded in the cosmic web \citep[e.g.,][]{Cantalupo2012a}.
Thus, the \hi\ absorption around the \MQ\ is expected to be suppressed.

We derive the radial profile of \hi\ overdensity $\delta_F$ around {\MQ} (hereafter \hi\ radial profile).  
Within the BQ1 region, 
we use the \hi\ tomographic map 
to calculate spherically averaged $\delta_F$ 
as a function of the 3D distance from {\MQ}, which is  
defined as 
\begin{eqnarray}
    R_{\rm 3D} \equiv \sqrt{d_{\rm RA}^2 +d_{\rm Dec}^2 + d_{z}^2}.
    \label{eq: 3D distance}
\end{eqnarray}
$d_{\rm RA}$, $d_{\rm Dec}$, and $d_{z}$ are the co-moving distances from {\MQ}  
under the assumption that the \hi\ absorbers have 
zero peculiar velocities relative to {\MQ}.  

To estimate the uncertainties of the spherically averaged $\delta_{F}$ values,  
we create mock Ly$\alpha$ forest transmission data 
by adding noise 
to the obtained $F(z)$ data 
based on the uncertainties of $F(z)$ estimated in Section \ref{sec: continuum estimation}, 
and calculate the $\delta_F$ values for the mock data along the $17$ sightlines. 
We then obtain a mock \hi\ tomographic map, 
and compute spherically averaged $\delta_F$ values as a function of $R_{\rm 3D}$. 
We repeat this process 1000 times 
and obtain $68${\%} intervals as the $1\sigma$ confidence intervals. 
Note that the typical $1\sigma$ uncertainty of $\delta_F$ 
for a pixel in the \hi\ tomographic map is found to be about $0.08$. 

The \hi\ tomographic map allows us to obtain the \hi\ radial profile up to around $R_{\rm 3D} = 6$ pMpc, 
which is limited due to the size of our \hi\ tomographic map. 
To extend our measurements beyond $6$ pMpc, 
obtain spherically averaged $\delta_F$ values 
based on the $\delta_F$ measurements along the sightlines of the background quasars in the BQ2--3 regions. 
Thanks to the large field coverage, 
our \hi\ radial profile measurements probe up to about $100$ pMpc around {\MQ}. 

\begin{figure}[t]
 {\centering
 \includegraphics[width=1.0\hsize, clip, bb= 0 0 765 650, clip=true]{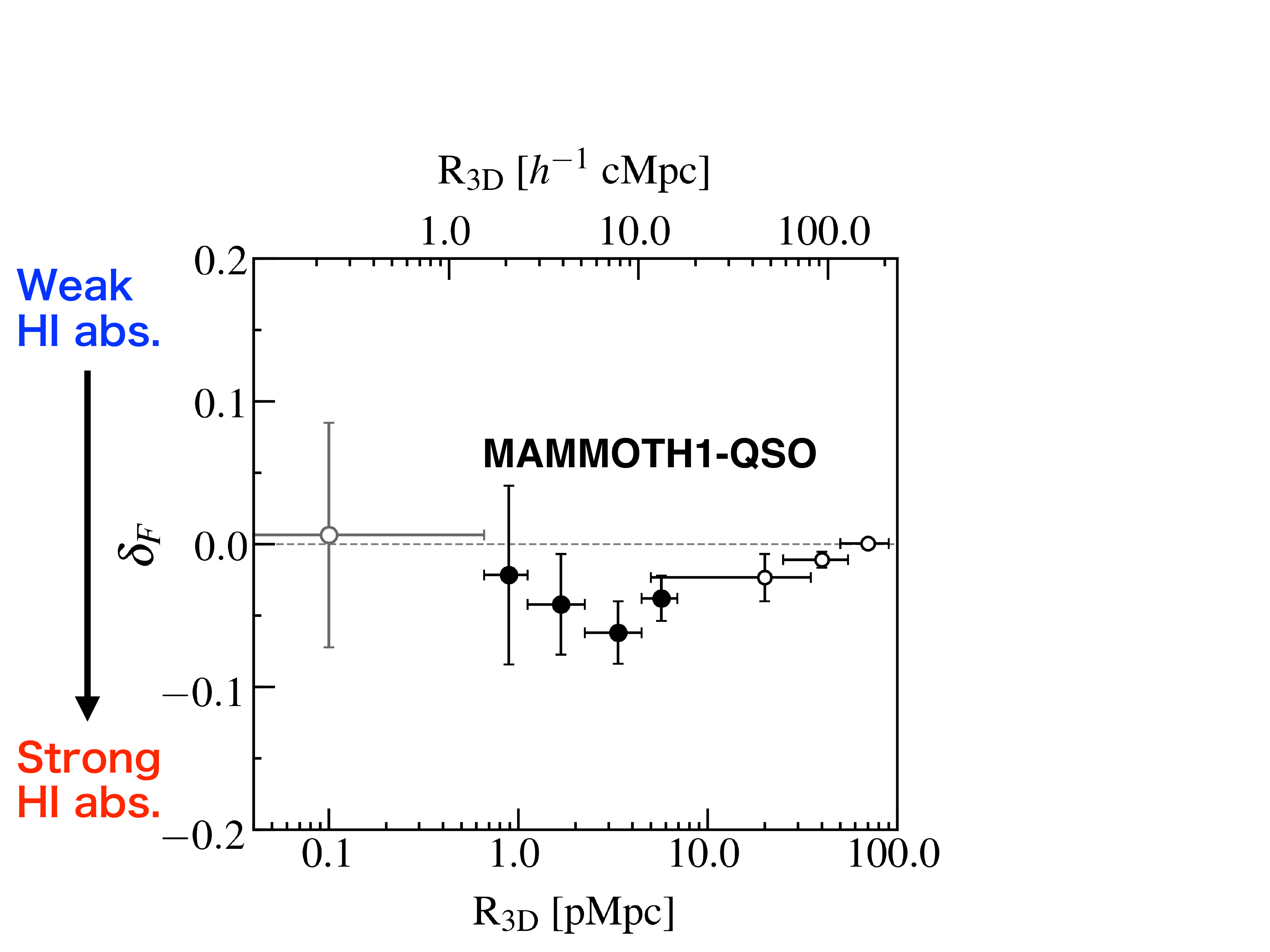}
 \caption{
 \hi\ radial profile around \MQ\ as a function of 3D distance $R_{\rm 3D}$ from it.
 The black filled (open) circles show our results obtained in the BQ1 region (BQ2 and BQ3 regions). 
 The black horizontal bars represent the measurement boundaries for the average. 
 The horizontal dotted line represents the cosmic average of the \hi\ absorption at $z=2.32$.
 The gray open circle presents the measurement of \hi\ tomography map pixel 
 where MAMMOTH1-QSO resides and
 whose spatial scale is below the spatial resolution
 which is shown as gray horizontal bars for reference.
 }
 }
 \label{fig: mq}
\end{figure}

Figure \ref{fig: mq} shows 
the obtained \hi\ radial profile around \MQ. 
We find that 
$\delta_{F}$ decreases (i.e., the strength of \hi\ absorption increases) 
with increasing $R_{\rm 3D}$ up to $\simeq 3$ pMpc 
from $\delta_{F} \simeq 0$ to $\delta_{F} = -0.06\pm0.02$, 
and  
$\delta_{F}$ slightly increases at larger distances.  
In other words, 
the \hi\ radial profile of \MQ\ 
shows a possible turnover at $R_{\rm 3D} \simeq 3 $ pMpc, 
indicating that  
\MQ\ resides in a volume with fairly weak \hi\ absorption.   
This tendency at small distances is opposite to that found for moderately bright galaxies at similar redshifts 
(e.g., \citealt{Rudie2012a}; \citealt{Rakic2012a}; \citealt{Turner2014a}); 
the \hi\ gas absorption around galaxies is stronger at smaller galactocentric radii. 
Our results may suggest that 
\MQ\ has a proximity zone where \hi\ gas is photo-ionized and \hi\ absorption is suppressed 
due to strong ionizing radiation from {\MQ}. 
In this picture, the ELAN around {\MQ} may be a photo-ionized cloud embedded in the cosmic web. 

We caution readers that 
our suggested picture is based on the tomographic map data as well as the background sources, which are partially sampling the space around \MQ\ (see Figures \ref{fig: sky} and \ref{fig: bkgls}).  
Although, we find a possible turnover in the \hi\ radial profile, 
the validity of this picture should be statistically tested with more background sources in future work.

Note that the \hi\ radial profile shows negative $\delta_F$ values at $R_{\rm 3D} \simeq 10$--$30$ pMpc, 
which is consistent with the detection of 
the strong \hi\ absorption group found by \cite{Cai2017b}. 
We also confirm that the $\delta_F$ values reach the cosmic average ($\delta_F = 0$) 
at a large scale of $R_{\rm 3D} \simeq 100$ pMpc. 

\subsubsection{Comparisons with Type-I Quasars} \label{sec: qpq6}
Since \MQ\ is categorized as a type-II quasar \citep{Cai2017a}, 
it is interesting to compare its \hi\ radial profile with those of type-I quasars. 

To make comparisons with type-I quasars, 
we select three type-I quasars 
(QSO1, QSO2, and QSO3) within our \hi\ tomographic map 
from the DR14Q catalog\footnote{\MQ\ is not observed in the eBOSS survey because of its faintness ($V=24.20$ mag).} 
and calculate spherically averaged \hi\ radial profiles around them 
in the same manner as that for \MQ.  
The basic properties of QSOs 1--3 are summarized in Table \ref{tab: neighboring qso} 
and their positions in the tomographic map 
are shown in Figure \ref{fig: HI3D}. 

\begin{figure}[t]
 {\centering
 \includegraphics[width=1.0\hsize, clip, bb= 0 0 750 650, clip=true]{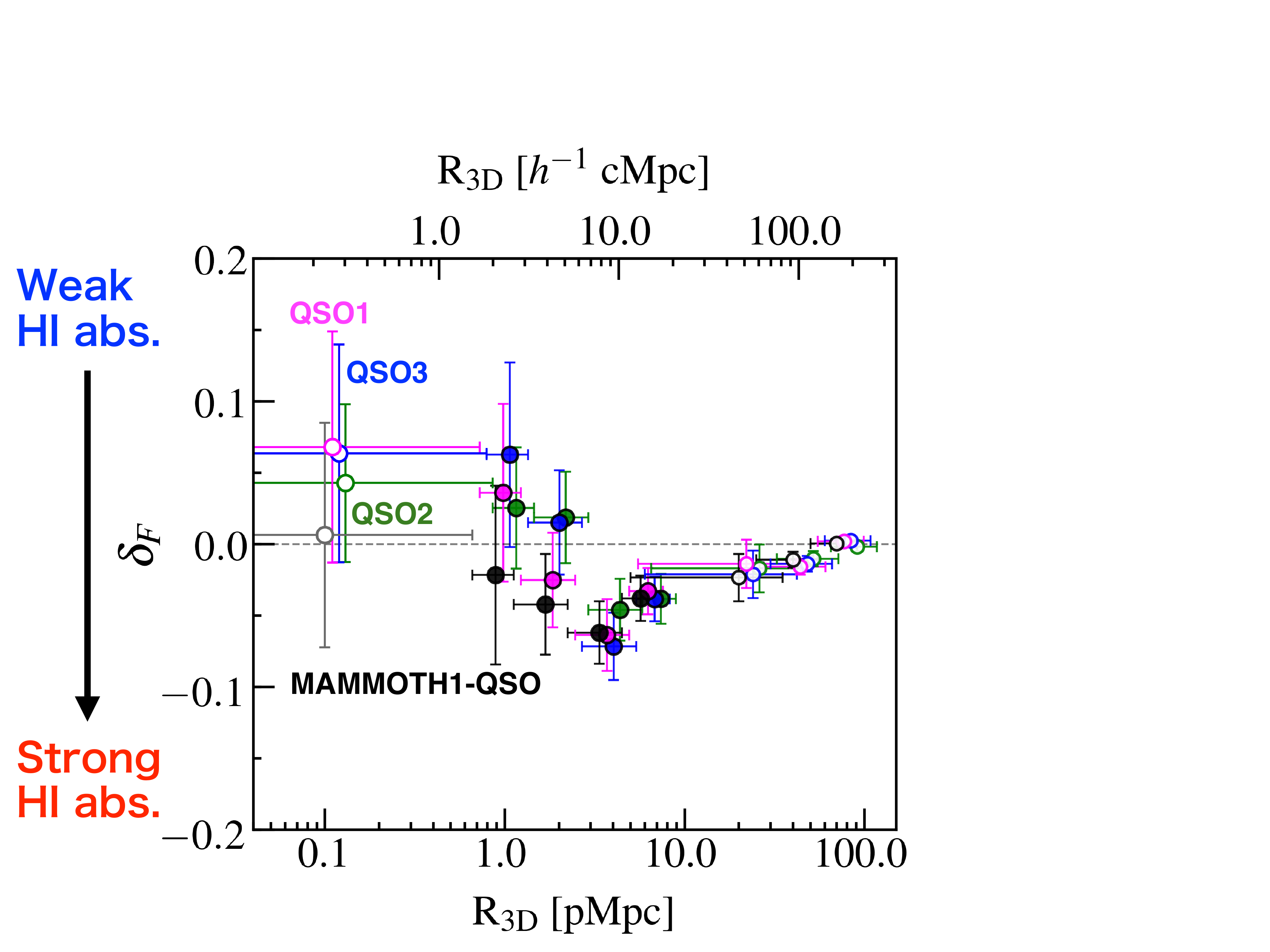}
 \caption{
 Same as Figure \ref{fig: mq}, but for 
 the three neighboring quasars (QSOs 1--3) as well as {\MQ}.
 The black, magenta, green, and blue circles represent \MQ, QSO1, QSO2, and QSO3, respectively. 
 The colored circles are slightly offset for clarity. 
 The innermost open circles present the measurements of \hi\ tomography map pixels 
 where QSOs 1--3 and {\MQ} reside and
 whose spatial scales are below the spatial resolution
 which is shown as the attached horizontal bars for reference.}
 }
 \label{fig: neighboring qsos}
\end{figure}

Figure \ref{fig: neighboring qsos} compares the \hi\ radial profiles of QSOs 1--3 
with that of \MQ.  
We find that 
their \hi\ radial profiles are similar to that of \MQ\ across $100$ pMpc, 
showing a common turnover at $R_{\rm 3D} \simeq 3 $ pMpc.
We should be cautious of the partial sampling around QSOs 1--3 (see Section \ref{sec: mq})
This result may indicate that 
spherically averaged \hi\ gas distributions around type-I and type-II quasars are similar, 
which is compatible with the AGN unification model  \citep[e.g.,][]{Antonucci1993a,Elvis2000a}:  
type-I quasars can ionize gas preferentially in the line-of-sight direction,
while type-II quasars can ionize gas in the transverse directions rather than the line-of-sight direction.  

\begin{deluxetable}{cccccccc}[t]
\tablecolumns{6} 
\tablewidth{0pt}
\vspace{-0em}
\tablecaption{
Three neighboring quasars in the \hi\ tomographic map selected from the DR14Q catalog 
\label{tab: neighboring qso}}
\tablehead{
\colhead{Source}&     \colhead{R.A.} &    \colhead{Decl.} &      \colhead{$z_{\rm spec}$} &  \colhead{$g$} &   \colhead{Ref.}\tablenotemark{a}   \\ 
\colhead{ }     &  \colhead{(J2000)} &  \colhead{(J2000)} &                   \colhead{ } & \colhead{(AB)} &     \colhead{ }
}
\startdata 
QSO1 &  14\,41\,33.75 & +40\,01\,42.78 &  $2.306$\tablenotemark{b}  & 20.66 &  DR14Q\\ 
QSO2 &  14\,40\,49.14 & +39\,54\,07.51 &  $2.306$\tablenotemark{b}  & 21.02 &  DR14Q\\ 
QSO3 &  14\,41\,21.66 & +40\,02\,58.82 &  $2.305$\tablenotemark{b}  & 21.87 &  DR14Q\\
\enddata 
\centering
\tablenotetext{$a$}{DR14Q: \cite{Paris2018a} }
\tablenotetext{$b$}{Spectroscopic redshift determined by MgII emission \citep{Paris2018a}.
                    The redshift uncertainty is $\simeq 300$ km s$^{-1}$ \citep{Prochaska2013a}.}
\end{deluxetable} 

\begin{figure*}[t]
 {\centering
 \includegraphics[width=1.0\hsize, clip, bb= 0 50 1050 720, clip=true]{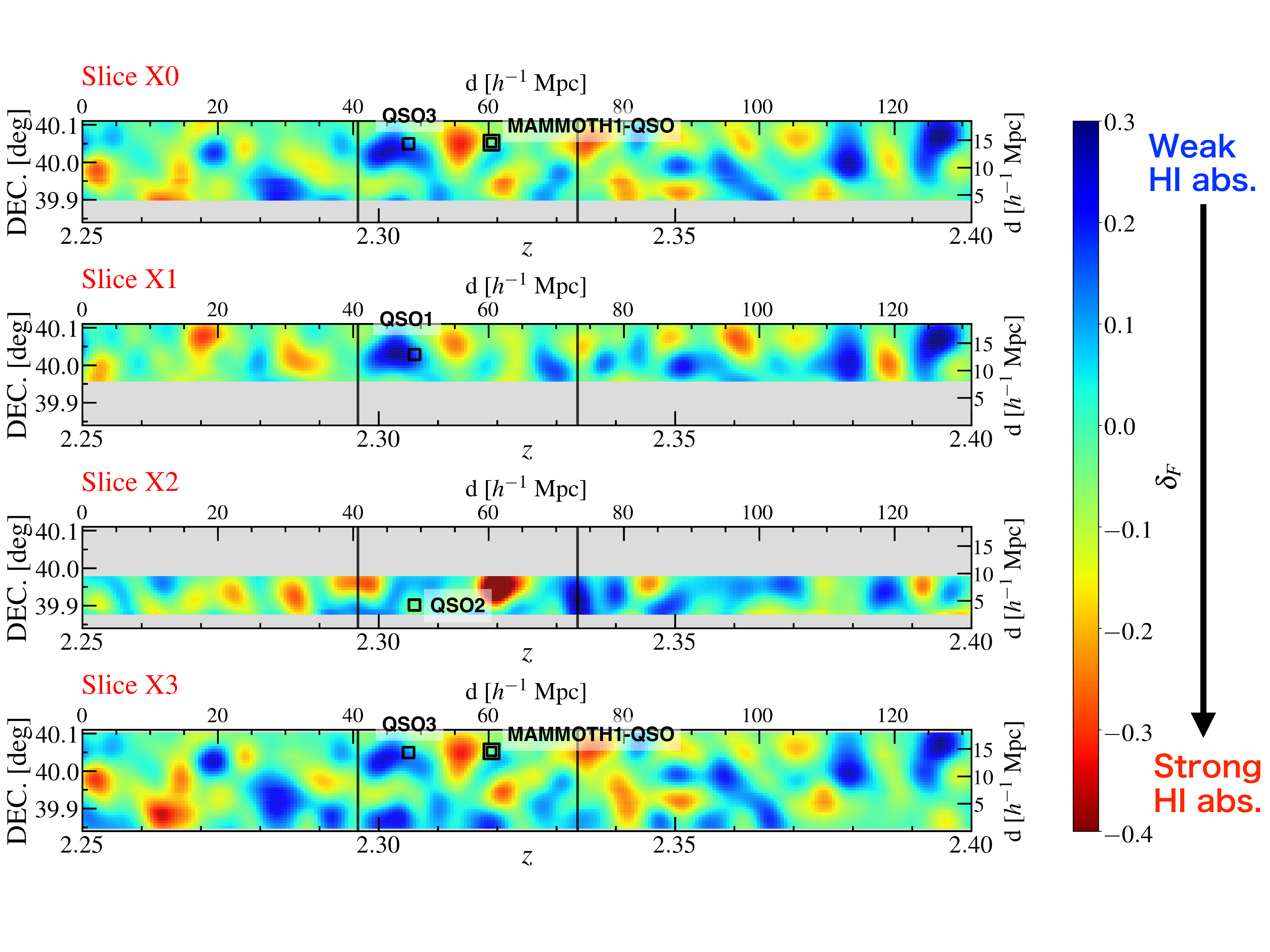}
 \caption{
 Projections of our \hi\ tomographic map 
 across $2.6 h^{-1}$ cMpc slices in the R.A. direction 
 centered at the positions of \MQ\ and QSOs 1--3 from top to bottom.  
 The projected R.A. ranges are shown in Figure \ref{fig: xy_lo}. 
 The color contours represent the \hi\ overdensity $\delta_{F}$ 
 such that negative values (red) correspond to higher overdensities. 
 The double square indicates the position of \MQ, 
 and the single squares show QSOs 1--3.  
 The two vertical lines at $z \simeq 2.30$ and $z \simeq 2.33$ 
 denote the redshift range where Ly$\alpha$ emission can be probed with NB403 
 used in \cite{Cai2017a, Cai2017b} (Section \ref{sec: LAE-HI overdensity}).
 }
 }
 \label{fig:fig4}
\end{figure*}

Figure \ref{fig:fig4} presents  
2D slices of the \hi\ tomographic map 
projected across $\Delta x = 2.6 h^{-1}$ cMpc along the $x$ (R.A.) direction 
around QSOs 1--3 and \MQ.   
The projected ranges along the $x$ direction for the four slices are shown in Figure \ref{fig: xy_lo}. 
In Figure \ref{fig:fig4}, 
we find that 
both QSOs 1--3 and \MQ\ 
are associated with or surrounded by 
\hi\ underdense regions with sizes of $\simeq 5-10\ h^{-1}$ cMpc, 
which would be created by strong photo-ionizing radiation from the quasars. 
Interestingly, these sizes are comparable to the estimated sizes of proximity zones of 
$z\sim2$ quasars \citep{DOdorico2008a}. 

Note that 
there is a representative of \hi\ absorption measurements
as a function of transverse distances to $z\sim2$ type-I quasars
performed by \cite{Prochaska2013a}, 
who have made use of a large ensemble of foreground/background quasar pairs 
based on the Quasar Probing Quasar survey (hereafter QPQ6).
We estimate \hi\ absorption 
around \MQ\ and QSOs 1--3 
with the same method as used in QPQ6, and 
present comparisons with the QPQ6 results in Appendix \ref{sec: APPA}.
We could not investigate \hi\ gas distributions
at small scales of $R_{\rm 3D} \lesssim 1$ pMpc where the QPQ6 study probes, 
because of the small number of background sightlines close to the quasars
and thus the large uncertainties.
The detailed comparison will be conducted with
future dense sampling of background sightlines 
at small scales of $R_{\rm 3D} \lesssim 1$ pMpc.

\subsection{LAE-HI overdensity} \label{sec: LAE-HI overdensity}
As described in Section \ref{sec:introduction}, 
the LAE overdense region has been found around \MQ\ 
in previous narrowband (NB403) imaging observations \citep{Cai2017b}. 
Since galaxies are good tracers of LSSs of the matter distribution in the universe, 
it is interesting to compare the spatial distribution of LAE overdensities with that of \hi\ overdensities. 

We compute the LAE overdensity based on the LAE sample 
constructed by \cite{Cai2017b}.\footnote{The detection limit of their observations 
corresponds to a \lya\ luminosity of $\sim 0.73 L_{\rm Ly\alpha}^{*}$, 
where $L_{\rm Ly\alpha}^{*} = 2.14 \times 10^{42}$ erg s$^{-1}$ is the characteristic \lya\ luminosity 
at $z=2.1$--$3.1$ \citep{Ciardullo2012a}.} 
The LAE overdensity $\delta_{\rm LAE}$ is defined as
\begin{eqnarray}
	\delta_{\rm LAE} \equiv \frac{ n_{\rm LAE} }{ \overline{n}_{\rm LAE} } -1,
	\label{eq:lae overdensity}
\end{eqnarray}
where $n_{\rm LAE}$ and $\overline{n}_{\rm LAE}$ are 
the LAE number density and its average, respectively,  
measured in a cylinder with a radius of $\langle d_{\perp }\rangle \simeq 2.6\ h^{-1}$ cMpc. 
For the cylinder length along the $z$ direction, 
we adopt a length of $\simeq 32.4 h^{-1}$ cMpc 
that corresponds to the redshift range where Ly$\alpha$ emission can be detected within the FWHM of NB403, 
i.e, $z = 2.30$--$2.33$.  
The obtained LAE overdensity is presented in Figure \ref{fig: sky}. 
We find two LAE LSSs whose density peaks are located at 
(R.A., Decl.) $=$ (14:41:27.12, +40:02:00.6) and (14:41:07840, +39:55:22.8). 

\begin{figure}[t]
 {\centering
 \includegraphics[width=1.1\hsize, clip, bb= 10 0 880 700, clip=true]{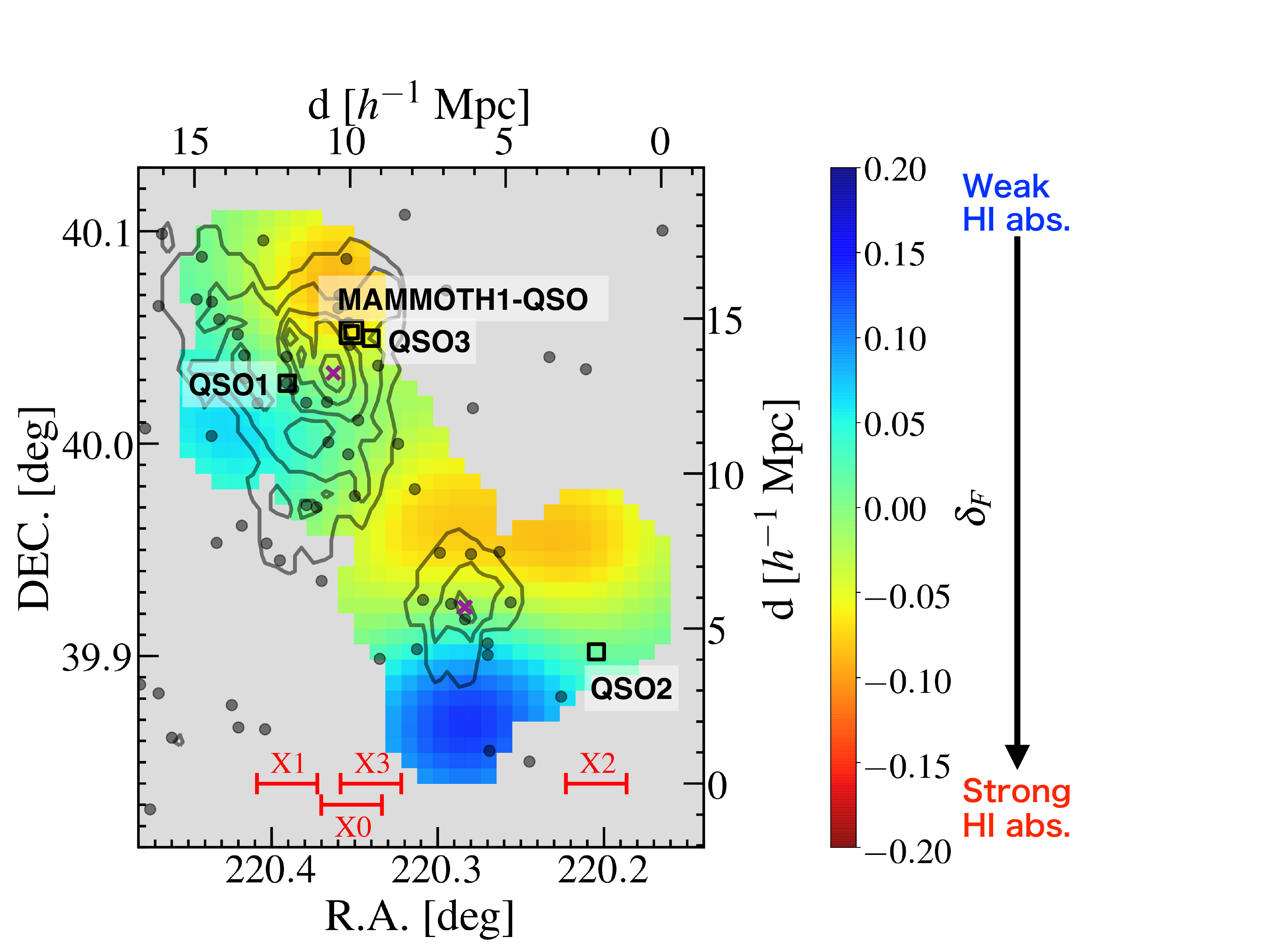}
 \caption{
 Projection of the \hi\ tomographic map over the redshift range of $z=2.30$--$2.33$. 
 The color contours represent the projected $\delta_{F}$.
 The gray dots are the LAEs found by \cite{Cai2017b} 
 and the gray contours represent the significances of the LAE overdensities from $2\sigma$ to $6\sigma$.
 The LAE distribution shows two density peaks that are marked with purple crosses. 
 The double square indicates the position of \MQ\  
 and the single squares are those of the three neighboring type-I quasars.
 The red bars X0--X3 correspond to the R.A. ranges adopted for the slices in Figure \ref{fig:fig4}. 
 }
 }
 \label{fig: xy_lo}
\end{figure}

In Figure \ref{fig: xy_lo}, 
we compare the sky distribution of the LAE overdensity 
with the projected \hi\ overdensity over the same redshift range of $z=2.30$--$2.33$ 
calculated from the \hi\ tomographic map. 
We find that the two LAE density peaks 
are spatially offset from the \hi\ density peaks by $\sim 3-5 h^{-1}$ cMpc.  
It is thought that galaxies are good tracers of underlying gas and dark matter distributions. 
However, since LAEs are star-forming galaxies and can emit ionizing photons, 
the \hi\ gas near the LAE density peaks would be relatively easily photo-ionized 
compared to LAE underdense regions.  
Such an anisotropic ionizing background radiation 
created by the density fluctuations of star-forming galaxies 
may cause this segregation between LAEs and \hi\ LSSs. 

Another interesting point in Figure \ref{fig: xy_lo}  
is that the two LAE overdense structures are bridged by 
one of the \hi\ overdense structures.  
This would be consistent with  
the picture that galaxy overdense structures are connected by the \hi\ cosmic web.

We also find that the position of \MM\ is located 
around the edges of the LAE overdense region and the \hi\ overdense region. 
In previous studies, 
LAEs with extended Ly$\alpha$ emission at similar redshifts 
tend to locate around the edges of galaxy overdense regions 
rather than the density peaks \citep{Badescu2017a, Mawatari2012a}. 
Our results are consistent with these previous results. 

\section{Conclusion} \label{sec:conclusion}
We have investigated the 3D distribution of IGM \hi\ gas around 
the ELAN of \MM\ at $z=2.3$. 
In a volume of $16 \times 19 \times 131 \ h^{-3} {\rm  cMpc}^{3}$ around \MQ,  
we have constructed an \hi\ tomographic map 
based on Ly$\alpha$ forest absorption 
detected in one eBOSS quasar and 16 Keck/LRIS galaxy spectra. 
By combining the \hi\ tomographic map results 
with \hi\ overdensity estimates 
based on background quasar spectra in the outer region, 
we have derived a spherically averaged \hi\ radial profile of \MQ\ 
over a wide range of scales from about $0.1$ pMpc to $100$ pMpc. 
Our results are summarized below.

\begin{enumerate}
\item 
The IGM \hi\ tomographic map reveals 
the existence of 
\hi\ overdense ($\delta_{F} \simeq -0.3-(-0.2)$) and underdense  ($\delta_{F} \simeq 0.3-0.2$) LSSs
with the size of $10-20\ h^{-1} {\rm cMpc}$ for the first time, 
indicating that the \hi\ gas distribution around \MQ\ is inhomogeneous.  
\item 
The \hi\ radial profile of \MQ\ 
has a possible turnover at $R_{\rm 3D} \simeq 3$ pMpc 
and may increase with decreasing $R_{\rm 3D}$, 
suggesting 
that \MQ\ may have 
a proximity zone 
where the quasar photo-ionizes the surrounding \hi\ gas and suppresses \hi\ absorption.
The \MM\ Ly$\alpha$ nebula is probably a photo-ionized cloud embedded in the cosmic web.
\item  
The \hi\ radial profile of {\MQ}, which is a type-II quasar, 
is similar to those of neighboring three type-I quasars at similar redshifts. 
This result suggests that 
spherically averaged \hi\ gas distributions around type-I and type-II quasars are similar, 
which is compatible with the AGN unification model. 
\item  
Based on a comparison between the \hi\ overdensity map and the distribution of LAEs around \MQ, 
we have found that their density peaks are spatially offset by about $3$--$5 h^{-1}$ cMpc. 
This spatial offset between the \hi\ and LAE LSSs 
may reflect anisotropic UV background radiation 
created by star-forming galaxy density fluctuations in this field.
\end{enumerate}

The connection between IGM \hi\ and galaxy formation in LSSs
can be systematically explored by a wide-field  spectroscopic survey of 
the Hobby-Eberly Telescope Dark Energy Experiment \citep[HETDEX; ][]{Hill2016a}.
The HETDEX survey will provide $10^6$ LAEs at $z\sim2-3$ over $400$ deg$^2$, and reveal a number of extended \lya\ nebulae and LSSs 
such as overdensities and filaments.
The LSS/IGM study with HETDEX will be complementary to 
the ongoing program of CLAMATO \citep{Lee2018a}, MAMMOTH \citep{Cai2016a}, and the planned program of gigantic IGM tomographic mapping with Subaru/PFS (K. Nagamine et al. in prep.).

\acknowledgments
We thank the anonymous referee for constructive comments and suggestions
that improved the clarity of the paper. 
We are grateful to Rieko Momose, Yujin Yang, 
Satoshi Kikuta, Siddhartha Gurung-Lopez, 
Marie W. Lau, and Koki Kakiichi
for their useful comments and discussions.
We also thank Percy Gomez and Nicholas McConnell
for their support of our Keck/LRIS observation.

The data presented herein were obtained at the W. M. Keck Observatory, which is operated as a scientific partnership among the California Institute of Technology, the University of California and the National Aeronautics and Space Administration. The Observatory was made possible by the generous financial support of the W. M. Keck Foundation.
The authors wish to recognize and acknowledge the very significant cultural role and reverence that the summit of Maunakea has always had within the indigenous Hawaiian community.  We are most fortunate to have the opportunity to conduct observations from this mountain.
This research has made use of the Keck Observatory Archive (KOA), which is operated by the W. M. Keck Observatory and the NASA Exoplanet Science Institute (NExScI), under contract with the National Aeronautics and Space Administration.
Funding for the Sloan Digital Sky Survey IV has been provided by the Alfred P. Sloan Foundation, the U.S. Department of Energy Office of Science, and the Participating Institutions. SDSS-IV acknowledges
support and resources from the Center for High-Performance Computing at
the University of Utah. The SDSS web site is www.sdss.org.
SDSS-IV is managed by the Astrophysical Research Consortium for the 
Participating Institutions of the SDSS Collaboration including the 
Brazilian Participation Group, the Carnegie Institution for Science, 
Carnegie Mellon University, the Chilean Participation Group, the French Participation Group, Harvard-Smithsonian Center for Astrophysics, 
Instituto de Astrof\'isica de Canarias, The Johns Hopkins University, Kavli Institute for the Physics and Mathematics of the Universe (IPMU) / 
University of Tokyo, the Korean Participation Group, Lawrence Berkeley National Laboratory, 
Leibniz Institut f\"ur Astrophysik Potsdam (AIP),  
Max-Planck-Institut f\"ur Astronomie (MPIA Heidelberg), 
Max-Planck-Institut f\"ur Astrophysik (MPA Garching), 
Max-Planck-Institut f\"ur Extraterrestrische Physik (MPE), 
National Astronomical Observatories of China, New Mexico State University, 
New York University, University of Notre Dame, 
Observat\'ario Nacional / MCTI, The Ohio State University, 
Pennsylvania State University, Shanghai Astronomical Observatory, 
United Kingdom Participation Group,
Universidad Nacional Aut\'onoma de M\'exico, University of Arizona, 
University of Colorado Boulder, University of Oxford, University of Portsmouth, 
University of Utah, University of Virginia, University of Washington, University of Wisconsin, 
Vanderbilt University, and Yale University.

This work is supported by World Premier International Research 
Center Initiative (WPI Initiative), MEXT, Japan, and 
KAKENHI (15H02064, 17H01110, and 17H01114) 
Grant-in-Aid for Scientific Research (A) 
through Japan Society for the Promotion of Science.
S.M. acknowledges support from the JSPS through the JSPS Research Fellowship for Young Scientists.
S.C. gratefully acknowledges support from Swiss National Science Foundation grant PP00P2\_163824.
Y.M. acknowledges support from the JSPS grants 17H04831, 17KK0098 and 19H00697.

\appendix
\section{\hi\ absorption as a function of transverse distances} \label{sec: APPA}
\begin{figure*}[t]
 \centering
 \includegraphics[width=1.0\hsize, clip, bb= 0 0 1000 700, clip=true]{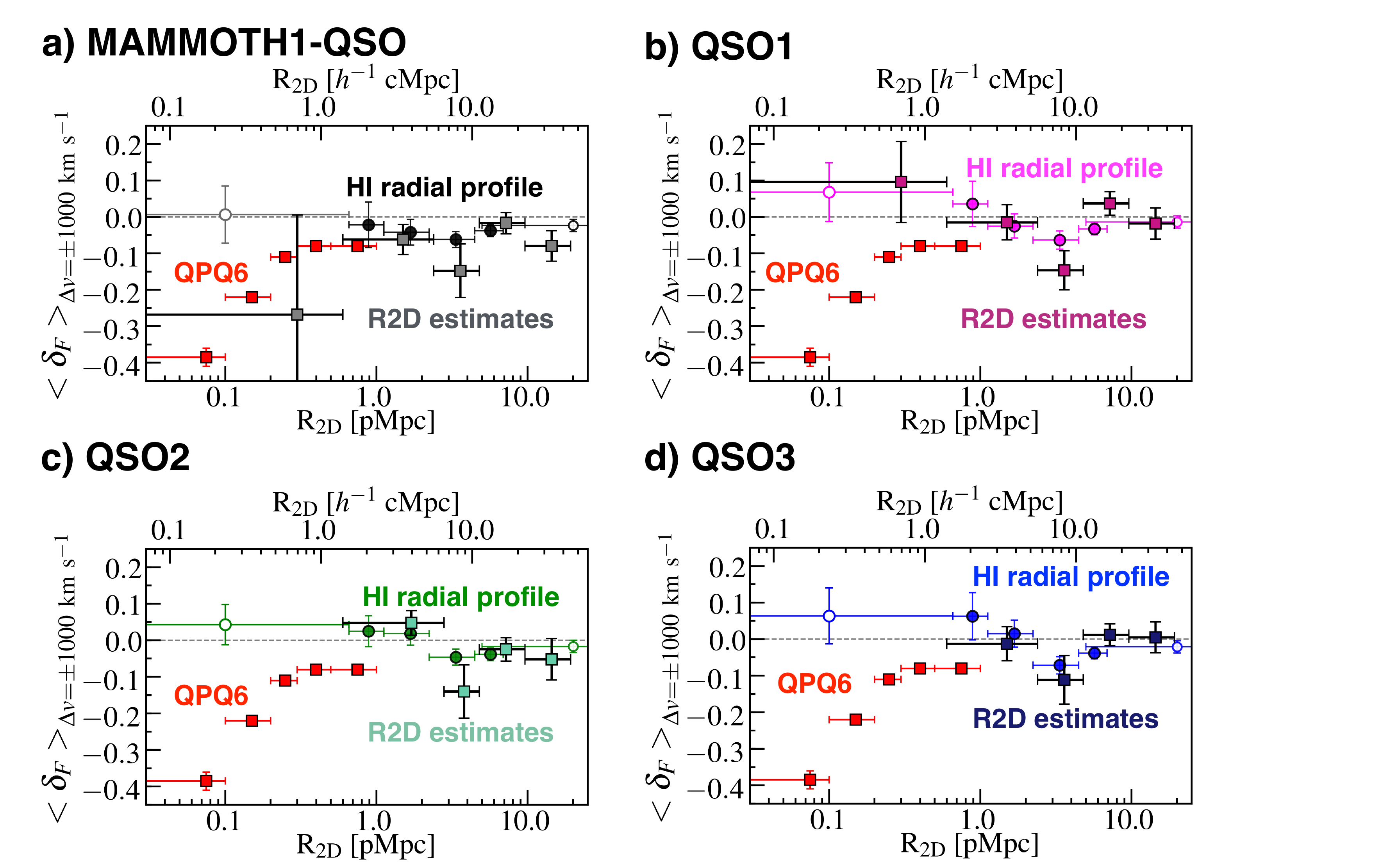}
 \caption{
 \hi\ absorption around \MQ\ and QSOs 1--3 estimated with the $R_{\rm 2D}$ measurements. 
 The (gray, deeppink, lightgreen, navy) squares are $R_{\rm 2D}$ measurements around (\MQ, 
 QSO1, QSO2, QSO3) in the BQ1 and BQ2 regions. Red squares are the results of QPQ6 
 \citep{Prochaska2013a}. 
 The (black, magenta, green, blue) circles are \hi\ radial profiles of (\MQ, QSO1, QSO2, QSO3)
 shown in Figures \ref{fig: mq} and \ref{fig: neighboring qsos}.
 The horizontal dotted line represents the cosmic average of the \hi\ absorption at $z=2.32$.
 }
 \label{fig: figAPPA1}
\end{figure*}

In this Appendix, we estimate \hi\ absorption around \MQ\ and QSOs 1--3, and compare with 
the QPQ6 results \citep{Prochaska2013a}.
We use our background source spectra, and apply the same method as used in QPQ6 (hereafter $R_{\rm 2D}$ measurements) in which we take average of \hi\ absorption in a $\pm1000$ km s$^{-1}$ velocity window around the QSOs as a function of transverse distances to the quasars ($R_{\rm 2D}$). 
The results for \MQ\ and QSOs 1--3 are shown in Figure \ref{fig: figAPPA1}. 
We find that our \hi\ absorption estimates are largely consistent with the QPQ6 result at $R_{\rm 2D} > 0.6$ pMpc ($=1.3$ \hmpc)  that corresponds to half the mean transverse sightline separation ${\langle d}_{\perp} \rangle$. We could not probe at $R_{\rm 2D} < 0.6$ pMpc because of the small number of background sightlines close to the quasars and thus the large uncertainties. 
In Figure \ref{fig: figAPPA1},
we also show our \hi\ radial profiles (Section \ref{sec: HI Tomography reconstruction}) to compare with the $R_{\rm 2D}$ measurements.
We should be cautious about this comparison,   
because the $R_{\rm 2D}$ measurements probe the \hi\ absorption
only in the transverse directions to the quasars while 
the \hi\ radial profiles are calculated with spherically averaged \hi\ overdensities, 
allowing us to probe the \hi\ radial profile averaged over all directions. 
Nevertheless, we find the possible turnovers at $\simeq3$ pMpc in both the measurements, which could support our argument of proximity zones in Section \ref{sec: HI Tomography reconstruction}.  
Since the data points of the \hi\ radial profiles below $0.6$ pMpc can be affected by interpolation in the tomographic reconstruction processes, these data are not well robust and we cannot directly compare with the QPQ6 results (See the caveat of the interpolation in Section \ref{sec: HI Tomography reconstruction}).
The detailed comparison will be conducted with future dense sampling of background sightlines.

\bibliography{citations.bib}

\begin{thebibliography}{}
\expandafter\ifx\csname natexlab\endcsname\relax\def\natexlab#1{#1}\fi
\providecommand{\url}[1]{\href{#1}{#1}}

\bibitem[{{Antonucci}(1993)}]{Antonucci1993a}
{Antonucci}, R. 1993, \araa, 31, 473

\bibitem[{{Becker} {et~al.}(2013){Becker}, {Hewett}, {Worseck}, \&
  {Prochaska}}]{Becker2013a}
{Becker}, G.~D., {Hewett}, P.~C., {Worseck}, G., \& {Prochaska}, J.~X. 2013,
  \mnras, 430, 2067

\bibitem[{{Berry} {et~al.}(2012){Berry}, {Gawiser}, {Guaita}, {Padilla},
  {Treister}, {Blanc}, {Ciardullo}, {Francke}, \& {Gronwall}}]{Berry2012a}
{Berry}, M., {Gawiser}, E., {Guaita}, L., {et~al.} 2012, \apj, 749, 4

\bibitem[{{Bertin} \& {Arnouts}(1996)}]{Bertin1996a}
{Bertin}, E., \& {Arnouts}, S. 1996, \aaps, 117, 393

\bibitem[{{Brammer} {et~al.}(2008){Brammer}, {van Dokkum}, \&
  {Coppi}}]{Brammer2008a}
{Brammer}, G.~B., {van Dokkum}, P.~G., \& {Coppi}, P. 2008, \apj, 686, 1503

\bibitem[{{Bruzual} \& {Charlot}(2003)}]{Bruzual2003a}
{Bruzual}, G., \& {Charlot}, S. 2003, \mnras, 344, 1000

\bibitem[{{B{\u{a}}descu} {et~al.}(2017){B{\u{a}}descu}, {Yang}, {Bertoldi},
  {Zabludoff}, {Karim}, \& {Magnelli}}]{Badescu2017a}
{B{\u{a}}descu}, T., {Yang}, Y., {Bertoldi}, F., {et~al.} 2017, \apj, 845, 172

\bibitem[{{Cai} {et~al.}(2016){Cai}, {Fan}, {Peirani}, {Bian}, {Frye},
  {McGreer}, {Prochaska}, {Lau}, {Tejos}, {Ho}, \& {Schneider}}]{Cai2016a}
{Cai}, Z., {Fan}, X., {Peirani}, S., {et~al.} 2016, \apj, 833, 135

\bibitem[{{Cai} {et~al.}(2017{\natexlab{a}}){Cai}, {Fan}, {Yang}, {Bian},
  {Prochaska}, {Zabludoff}, {McGreer}, {Zheng}, {Green}, {Cantalupo}, {Frye},
  {Hamden}, {Jiang}, {Kashikawa}, \& {Wang}}]{Cai2017a}
{Cai}, Z., {Fan}, X., {Yang}, Y., {et~al.} 2017{\natexlab{a}}, \apj, 837, 71

\bibitem[{{Cai} {et~al.}(2017{\natexlab{b}}){Cai}, {Fan}, {Bian}, {Zabludoff},
  {Yang}, {Prochaska}, {McGreer}, {Zheng}, {Kashikawa}, {Wang}, {Frye},
  {Green}, \& {Jiang}}]{Cai2017b}
{Cai}, Z., {Fan}, X., {Bian}, F., {et~al.} 2017{\natexlab{b}}, \apj, 839, 131

\bibitem[{{Cai} {et~al.}(2019){Cai}, {Cantalupo}, {Prochaska}, {Arrigoni
  Battaia}, {Burchett}, {Li}, {Chisholm}, {Bundy}, \& {Hennawi}}]{Cai2019a}
{Cai}, Z., {Cantalupo}, S., {Prochaska}, J.~X., {et~al.} 2019, \apjs, 245, 23

\bibitem[{{Calzetti} {et~al.}(2000){Calzetti}, {Armus}, {Bohlin}, {Kinney},
  {Koornneef}, \& {Storchi-Bergmann}}]{Calzetti2000a}
{Calzetti}, D., {Armus}, L., {Bohlin}, R.~C., {et~al.} 2000, \apj, 533, 682

\bibitem[{{Cantalupo} {et~al.}(2014){Cantalupo}, {Arrigoni-Battaia},
  {Prochaska}, {Hennawi}, \& {Madau}}]{Cantalupo2014a}
{Cantalupo}, S., {Arrigoni-Battaia}, F., {Prochaska}, J.~X., {Hennawi}, J.~F.,
  \& {Madau}, P. 2014, \nat, 506, 63

\bibitem[{{Cantalupo} {et~al.}(2012){Cantalupo}, {Lilly}, \&
  {Haehnelt}}]{Cantalupo2012a}
{Cantalupo}, S., {Lilly}, S.~J., \& {Haehnelt}, M.~G. 2012, \mnras, 425, 1992

\bibitem[{{Casali} {et~al.}(2007){Casali}, {Adamson}, {Alves de Oliveira},
  {Almaini}, {Burch}, {Chuter}, {Elliot}, {Folger}, {Foucaud}, {Hambly},
  {Hastie}, {Henry}, {Hirst}, {Irwin}, {Ives}, {Lawrence}, {Laidlaw}, {Lee},
  {Lewis}, {Lunney}, {McLay}, {Montgomery}, {Pickup}, {Read}, {Rees}, {Robson},
  {Sekiguchi}, {Vick}, {Warren}, \& {Woodward}}]{Casali2007a}
{Casali}, M., {Adamson}, A., {Alves de Oliveira}, C., {et~al.} 2007, \aap, 467,
  777

\bibitem[{{Chabrier}(2003)}]{Chabrier2003a}
{Chabrier}, G. 2003, \pasp, 115, 763

\bibitem[{{Ciardullo} {et~al.}(2012){Ciardullo}, {Gronwall}, {Wolf},
  {McCathran}, {Bond}, {Gawiser}, {Guaita}, {Feldmeier}, {Treister}, {Padilla},
  {Francke}, {Matkovi{\'c}}, {Altmann}, \& {Herrera}}]{Ciardullo2012a}
{Ciardullo}, R., {Gronwall}, C., {Wolf}, C., {et~al.} 2012, \apj, 744, 110

\bibitem[{{D'Odorico} {et~al.}(2008){D'Odorico}, {Bruscoli}, {Saitta},
  {Fontanot}, {Viel}, {Cristiani}, \& {Monaco}}]{DOdorico2008a}
{D'Odorico}, V., {Bruscoli}, M., {Saitta}, F., {et~al.} 2008, \mnras, 389, 1727

\bibitem[{{Elvis}(2000)}]{Elvis2000a}
{Elvis}, M. 2000, \apj, 545, 63

\bibitem[{{Faucher-Gigu{\`e}re} {et~al.}(2008){Faucher-Gigu{\`e}re},
  {Prochaska}, {Lidz}, {Hernquist}, \& {Zaldarriaga}}]{Faucher-Giguere2008a}
{Faucher-Gigu{\`e}re}, C.-A., {Prochaska}, J.~X., {Lidz}, A., {Hernquist}, L.,
  \& {Zaldarriaga}, M. 2008, \apj, 681, 831

\bibitem[{{Hayashino} {et~al.}(2019){Hayashino}, {Inoue}, {Kousai},
  {Kashikawa}, {Mawatari}, {Matsuda}, {Tejos}, {Prochaska}, {Iwata}, {Noll},
  {Burgarella}, {Yamada}, \& {Akiyama}}]{Hayashino2019a}
{Hayashino}, T., {Inoue}, A.~K., {Kousai}, K., {et~al.} 2019, \mnras, 484, 5868

\bibitem[{{Hill} \& {HETDEX Consortium}(2016)}]{Hill2016a}
{Hill}, G.~J., \& {HETDEX Consortium}. 2016, in Astronomical Society of the
  Pacific Conference Series, Vol. 507, Multi-Object Spectroscopy in the Next
  Decade: Big Questions, Large Surveys, and Wide Fields, ed. I.~{Skillen},
  M.~{Barcells}, \& S.~{Trager}, 393

\bibitem[{{Hinshaw} {et~al.}(2013){Hinshaw}, {Larson}, {Komatsu}, {Spergel},
  {Bennett}, {Dunkley}, {Nolta}, {Halpern}, {Hill}, {Odegard}, {Page}, {Smith},
  {Weiland}, {Gold}, {Jarosik}, {Kogut}, {Limon}, {Meyer}, {Tucker}, {Wollack},
  \& {Wright}}]{Hinshaw2013a}
{Hinshaw}, G., {Larson}, D., {Komatsu}, E., {et~al.} 2013, \apjs, 208, 19

\bibitem[{{Inoue} {et~al.}(2014){Inoue}, {Shimizu}, {Iwata}, \&
  {Tanaka}}]{Inoue2014a}
{Inoue}, A.~K., {Shimizu}, I., {Iwata}, I., \& {Tanaka}, M. 2014, \mnras, 442,
  1805

\bibitem[{{Kikuta} {et~al.}(2019){Kikuta}, {Matsuda}, {Cen}, {Steidel}, {Yagi},
  {Hayashino}, {Imanishi}, {Komiyama}, {Momose}, \& {Saito}}]{Kikuta2019}
{Kikuta}, S., {Matsuda}, Y., {Cen}, R., {et~al.} 2019, \pasj, 71, L2

\bibitem[{{Lee} {et~al.}(2014{\natexlab{a}}){Lee}, {Hennawi}, {White}, {Croft},
  \& {Ozbek}}]{Lee2014a}
{Lee}, K.-G., {Hennawi}, J.~F., {White}, M., {Croft}, R.~A.~C., \& {Ozbek}, M.
  2014{\natexlab{a}}, \apj, 788, 49

\bibitem[{{Lee} {et~al.}(2012){Lee}, {Suzuki}, \& {Spergel}}]{Lee2012a}
{Lee}, K.-G., {Suzuki}, N., \& {Spergel}, D.~N. 2012, \aj, 143, 51

\bibitem[{{Lee} {et~al.}(2013){Lee}, {Bailey}, {Bartsch}, {Carithers},
  {Dawson}, {Kirkby}, {Lundgren}, {Margala}, {Palanque-Delabrouille}, {Pieri},
  {Schlegel}, {Weinberg}, {Y{\`e}che}, {Aubourg}, {Bautista}, {Bizyaev},
  {Blomqvist}, {Bolton}, {Borde}, {Brewington}, {Busca}, {Croft}, {Delubac},
  {Ebelke}, {Eisenstein}, {Font-Ribera}, {Ge}, {Hamilton}, {Hennawi}, {Ho},
  {Honscheid}, {Le Goff}, {Malanushenko}, {Malanushenko}, {Miralda-Escud{\'e}},
  {Myers}, {Noterdaeme}, {Oravetz}, {Pan}, {P{\^a}ris}, {Petitjean}, {Rich},
  {Rollinde}, {Ross}, {Rossi}, {Schneider}, {Simmons}, {Snedden}, {Slosar},
  {Spergel}, {Suzuki}, {Viel}, \& {Weaver}}]{Lee2013a}
{Lee}, K.-G., {Bailey}, S., {Bartsch}, L.~E., {et~al.} 2013, \aj, 145, 69

\bibitem[{{Lee} {et~al.}(2014{\natexlab{b}}){Lee}, {Hennawi}, {Stark},
  {Prochaska}, {White}, {Schlegel}, {Eilers}, {Arinyo-i-Prats}, {Suzuki},
  {Croft}, {Caputi}, {Cassata}, {Ilbert}, {Garilli}, {Koekemoer}, {Le Brun},
  {Le F{\`e}vre}, {Maccagni}, {Nugent}, {Taniguchi}, {Tasca}, {Tresse},
  {Zamorani}, \& {Zucca}}]{Lee2014b}
{Lee}, K.-G., {Hennawi}, J.~F., {Stark}, C., {et~al.} 2014{\natexlab{b}},
  \apjl, 795, L12

\bibitem[{{Lee} {et~al.}(2016){Lee}, {Hennawi}, {White}, {Prochaska},
  {Font-Ribera}, {Schlegel}, {Rich}, {Suzuki}, {Stark}, {Le F{\`e}vre},
  {Nugent}, {Salvato}, \& {Zamorani}}]{Lee2016a}
{Lee}, K.-G., {Hennawi}, J.~F., {White}, M., {et~al.} 2016, \apj, 817, 160

\bibitem[{{Lee} {et~al.}(2018){Lee}, {Krolewski}, {White}, {Schlegel},
  {Nugent}, {Hennawi}, {M{\"u}ller}, {Pan}, {Prochaska}, {Font-Ribera},
  {Suzuki}, {Glazebrook}, {Kacprzak}, {Kartaltepe}, {Koekemoer}, {Le
  F{\`e}vre}, {Lemaux}, {Maier}, {Nanayakkara}, {Rich}, {Sanders}, {Salvato},
  {Tasca}, \& {Tran}}]{Lee2018a}
{Lee}, K.-G., {Krolewski}, A., {White}, M., {et~al.} 2018, \apjs, 237, 31

\bibitem[{{Mawatari} {et~al.}(2012){Mawatari}, {Yamada}, {Nakamura},
  {Hayashino}, \& {Matsuda}}]{Mawatari2012a}
{Mawatari}, K., {Yamada}, T., {Nakamura}, Y., {Hayashino}, T., \& {Matsuda}, Y.
  2012, \apj, 759, 133

\bibitem[{{Mawatari} {et~al.}(2017){Mawatari}, {Inoue}, {Yamada}, {Hayashino},
  {Otsuka}, {Matsuda}, {Umehata}, {Ouchi}, \& {Mukae}}]{Mawatari2017a}
{Mawatari}, K., {Inoue}, A.~K., {Yamada}, T., {et~al.} 2017, \mnras, 467, 3951

\bibitem[{{Mukae} {et~al.}(2017){Mukae}, {Ouchi}, {Kakiichi}, {Suzuki}, {Ono},
  {Cai}, {Inoue}, {Chiang}, {Shibuya}, \& {Matsuda}}]{Mukae2017a}
{Mukae}, S., {Ouchi}, M., {Kakiichi}, K., {et~al.} 2017, \apj, 835, 281

\bibitem[{{Myers} {et~al.}(2015){Myers}, {Palanque-Delabrouille}, {Prakash},
  {P{\^a}ris}, {Yeche}, {Dawson}, {Bovy}, {Lang}, {Schlegel}, {Newman},
  {Petitjean}, {Kneib}, {Laurent}, {Percival}, {Ross}, {Seo}, {Tinker},
  {Armengaud}, {Brownstein}, {Burtin}, {Cai}, {Comparat}, {Kasliwal},
  {Kulkarni}, {Laher}, {Levitan}, {McBride}, {McGreer}, {Miller}, {Nugent},
  {Ofek}, {Rossi}, {Ruan}, {Schneider}, {Sesar}, {Streblyanska}, \&
  {Surace}}]{Myers2015a}
{Myers}, A.~D., {Palanque-Delabrouille}, N., {Prakash}, A., {et~al.} 2015,
  \apjs, 221, 27

\bibitem[{{Noterdaeme} {et~al.}(2012){Noterdaeme}, {Petitjean}, {Carithers},
  {P{\^a}ris}, {Font-Ribera}, {Bailey}, {Aubourg}, {Bizyaev}, {Ebelke},
  {Finley}, {Ge}, {Malanushenko}, {Malanushenko}, {Miralda-Escud{\'e}},
  {Myers}, {Oravetz}, {Pan}, {Pieri}, {Ross}, {Schneider}, {Simmons}, \&
  {York}}]{Noterdaeme2012a}
{Noterdaeme}, P., {Petitjean}, P., {Carithers}, W.~C., {et~al.} 2012, \aap,
  547, L1

\bibitem[{{Oke} \& {Gunn}(1983)}]{Oke1983a}
{Oke}, J.~B., \& {Gunn}, J.~E. 1983, \apj, 266, 713

\bibitem[{{Oke} {et~al.}(1995){Oke}, {Cohen}, {Carr}, {Cromer}, {Dingizian},
  {Harris}, {Labrecque}, {Lucinio}, {Schaal}, {Epps}, \& {Miller}}]{Oke1995a}
{Oke}, J.~B., {Cohen}, J.~G., {Carr}, M., {et~al.} 1995, \pasp, 107, 375

\bibitem[{{P{\^a}ris} {et~al.}(2017){P{\^a}ris}, {Petitjean}, {Ross}, {Myers},
  {Aubourg}, {Streblyanska}, {Bailey}, {Armengaud}, {Palanque-Delabrouille},
  {Y{\`e}che}, {Hamann}, {Strauss}, {Albareti}, {Bovy}, {Bizyaev}, {Niel
  Brandt}, {Brusa}, {Buchner}, {Comparat}, {Croft}, {Dwelly}, {Fan},
  {Font-Ribera}, {Ge}, {Georgakakis}, {Hall}, {Jiang}, {Kinemuchi},
  {Malanushenko}, {Malanushenko}, {McMahon}, {Menzel}, {Merloni}, {Nandra},
  {Noterdaeme}, {Oravetz}, {Pan}, {Pieri}, {Prada}, {Salvato}, {Schlegel},
  {Schneider}, {Simmons}, {Viel}, {Weinberg}, \& {Zhu}}]{Paris2017a}
{P{\^a}ris}, I., {Petitjean}, P., {Ross}, N.~P., {et~al.} 2017, \aap, 597, A79

\bibitem[{{P{\^a}ris} {et~al.}(2018){P{\^a}ris}, {Petitjean}, {Aubourg},
  {Myers}, {Streblyanska}, {Lyke}, {Anderson}, {Armengaud}, {Bautista},
  {Blanton}, {Blomqvist}, {Brinkmann}, {Brownstein}, {Brand t}, {Burtin},
  {Dawson}, {de la Torre}, {Georgakakis}, {Gil-Mar{\'\i}n}, {Green}, {Hall},
  {Kneib}, {LaMassa}, {Le Goff}, {MacLeod}, {Mariappan}, {McGreer}, {Merloni},
  {Noterdaeme}, {Palanque-Delabrouille}, {Percival}, {Ross}, {Rossi},
  {Schneider}, {Seo}, {Tojeiro}, {Weaver}, {Weijmans}, {Y{\`e}che}, {Zarrouk},
  \& {Zhao}}]{Paris2018a}
{P{\^a}ris}, I., {Petitjean}, P., {Aubourg}, {\'E}., {et~al.} 2018, \aap, 613,
  A51

\bibitem[{{Pedichini} {et~al.}(2003){Pedichini}, {Giallongo}, {Ragazzoni}, {Di
  Paola}, {Fontana}, {Speziali}, {Farinato}, {Baruffolo}, {Magagna},
  {Diolaiti}, {Pasian}, {Smareglia}, {Anaclerio}, {Gallieni}, \&
  {Lazzarini}}]{Pedichini2003a}
{Pedichini}, F., {Giallongo}, E., {Ragazzoni}, R., {et~al.} 2003, in \procspie,
  Vol. 4841, Instrument Design and Performance for Optical/Infrared
  Ground-based Telescopes, ed. M.~{Iye} \& A.~F.~M. {Moorwood}, 815--826

\bibitem[{{Pettini} {et~al.}(2001){Pettini}, {Shapley}, {Steidel}, {Cuby},
  {Dickinson}, {Moorwood}, {Adelberger}, \& {Giavalisco}}]{Pettini2001a}
{Pettini}, M., {Shapley}, A.~E., {Steidel}, C.~C., {et~al.} 2001, \apj, 554,
  981

\bibitem[{{Pettini} {et~al.}(2000){Pettini}, {Steidel}, {Adelberger},
  {Dickinson}, \& {Giavalisco}}]{Pettini2000a}
{Pettini}, M., {Steidel}, C.~C., {Adelberger}, K.~L., {Dickinson}, M., \&
  {Giavalisco}, M. 2000, \apj, 528, 96

\bibitem[{{Prochaska} {et~al.}(2013){Prochaska}, {Hennawi}, {Lee}, {Cantalupo},
  {Bovy}, {Djorgovski}, {Ellison}, {Lau}, {Martin}, {Myers}, {Rubin}, \&
  {Simcoe}}]{Prochaska2013a}
{Prochaska}, J.~X., {Hennawi}, J.~F., {Lee}, K.-G., {et~al.} 2013, \apj, 776,
  136

\bibitem[{{Rakic} {et~al.}(2012){Rakic}, {Schaye}, {Steidel}, \&
  {Rudie}}]{Rakic2012a}
{Rakic}, O., {Schaye}, J., {Steidel}, C.~C., \& {Rudie}, G.~C. 2012, \apj, 751,
  94

\bibitem[{{Rudie} {et~al.}(2012){Rudie}, {Steidel}, {Trainor}, {Rakic},
  {Bogosavljevi{\'c}}, {Pettini}, {Reddy}, {Shapley}, {Erb}, \&
  {Law}}]{Rudie2012a}
{Rudie}, G.~C., {Steidel}, C.~C., {Trainor}, R.~F., {et~al.} 2012, \apj, 750,
  67

\bibitem[{{Shapley} {et~al.}(2003){Shapley}, {Steidel}, {Pettini}, \&
  {Adelberger}}]{Shapley2003a}
{Shapley}, A.~E., {Steidel}, C.~C., {Pettini}, M., \& {Adelberger}, K.~L. 2003,
  \apj, 588, 65

\bibitem[{{Stark} {et~al.}(2015){Stark}, {White}, {Lee}, \&
  {Hennawi}}]{Stark2015a}
{Stark}, C.~W., {White}, M., {Lee}, K.-G., \& {Hennawi}, J.~F. 2015, \mnras,
  453, 311

\bibitem[{{Suzuki} {et~al.}(2005){Suzuki}, {Tytler}, {Kirkman}, {O'Meara}, \&
  {Lubin}}]{Suzuki2005a}
{Suzuki}, N., {Tytler}, D., {Kirkman}, D., {O'Meara}, J.~M., \& {Lubin}, D.
  2005, \apj, 618, 592

\bibitem[{{Turner} {et~al.}(2014){Turner}, {Schaye}, {Steidel}, {Rudie}, \&
  {Strom}}]{Turner2014a}
{Turner}, M.~L., {Schaye}, J., {Steidel}, C.~C., {Rudie}, G.~C., \& {Strom},
  A.~L. 2014, \mnras, 445, 794

\end{thebibliography}

\end{document}